\documentclass{article}

\usepackage{arxiv}

\usepackage[utf8]{inputenc} 
\usepackage[T1]{fontenc}    
\usepackage{hyperref}       
\usepackage{url}            
\usepackage{booktabs}       
\usepackage{amsfonts}       
\usepackage{nicefrac}       
\usepackage{microtype}      
\usepackage{lipsum}

\usepackage{natbib}
\usepackage{graphicx}
\usepackage{subcaption}
\usepackage{amssymb}
\usepackage{soul}
\usepackage{textcomp}
\usepackage{xcolor}
\usepackage{booktabs}
\usepackage{multirow}
\usepackage{setspace}
\title{A comparison of oscillatory characteristics in covert speech and speech perception}

\author{
  Jae Moon\\
  Institue of Biomedical Engineering\\
  University of Toronto\\
  Bloorview Research Institute \\
  Holland Bloorview Kid's Rehabilitation Hospital\\
  \texttt{jae.moon@mail.utoronto.ca} \\
  \And
  Silvia Orlandi \\
  Bloorview Research Institute \\
  Holland Bloorview Kid's Rehabilitation Hospital\\
  \texttt{sorlandi@hollandbloorview.ca} \\
     \And
 Tom Chau \\
  Institue of Biomedical Engineering\\
  University of Toronto\\
  Bloorview Research Institute \\
  Holland Bloorview Kid's Rehabilitation Hospital\\
  \texttt{tom.chau@utoronto.ca} \\
}

\begin{document}
\maketitle



\begin{abstract}

Covert speech, the silent production of words in the mind, has been studied increasingly to understand and decode thoughts. This task has often been compared to speech perception as it brings about similar topographical activation patterns in common brain areas. In studies of speech comprehension, neural oscillations are thought to play a key role in the sampling of speech at varying temporal scales. However, very little is known about the role of oscillations in covert speech. In this study, we aimed to determine to what extent each oscillatory frequency band is used to process words in covert speech and speech perception tasks. Secondly, we asked whether the $\theta$ and $\gamma$ activity in the two tasks are related through phase-amplitude coupling (PAC). First, continuous wavelet transform was performed on epoched signals and subsequently two-tailed t-tests between two classes were conducted to determine statistical distinctions in frequency and time. While the perception task dynamically uses all frequencies with more prominent $\theta$ and $\gamma$ activity, the covert task favoured higher frequencies with significantly higher $\gamma$ activity than perception. Moreover, the perception condition produced significant $\theta$-$\gamma$ PAC suggesting a linkage of syllabic and phonological sampling. Although this was found to be suppressed in the covert condition, we found significant pseudo-coupling between perception $\theta$ and covert speech $\gamma$. We report that covert speech processing is largely conducted by higher frequencies, and that the $\gamma$- and $\theta$-bands may function similarly and differently across tasks, respectively. This study is the first to characterize covert speech in terms of neural oscillatory engagement. Future studies are directed to explore oscillatory characteristics and inter-task relationships with a more diverse vocabulary.


\end{abstract}





\section{Introduction}
Covert speech (CS), the silent production of words in one’s mind, is a fundamental trait in mental cognition \citep{Alderson-Day2012,Perrone-Bertolotti2014a}. It is referred to as a linguistic form of thought and linked to a wide range of neurocognitive functions, such as reading, writing, planning, and memory \citep{Alderson-Day2018,Morin2011,Morin2018}. Due to its ubiquity, many researchers in the realm of brain-computer interfaces (BCIs) have been assessing this task to restore speech in motor-impaired individuals by decoding thoughts \citep{DaSalla2009,Idrees2016,Deng2010}. However, CS BCIs are notorious for their difficulty in training, often requiring individuals to mentally rehearse each speech item numerous times for the system to learn a reliable control signal. Fortunately, research in neurolinguistics has linked CS with the task of speech perception (SP) due to theories and evidence describing the parallel nature of top-down and bottom-up language pathways. For instance, functional magnetic resonance imaging (fMRI) studies have revealed that CS and SP activate common brain regions along the linguistic processing pathway \citep{Okada2006,Shergill2002,Skipper2005,VandeVen2009,Venezia2016}, and time-domain methods report that the pattern of activation in these brain regions is similar \citep{Tian2010,Tian2012}. The results of these studies can suggest that CS signals could be modeled based on SP signals and that a CS BCI can be trained through the passive perception of speech. Thus, being able to model CS from SP would help hurdle the fatigue barrier in CS BCIs and enhance their translational potential. However, in order to achieve this modelling, one must understand how CS and SP tasks comparatively utilize neural oscillations, the primary mechanism of information transmission in the brain \citep{Buzsaki2004,Buzsaki2004a,Morillon2015,Luo2007,Ding2017}.



Numerous studies support a bi-directional linkage between perception and production systems of speech \citep{Buchsbaum2001,Hickok2007,Hickok2004,Poeppel2014,Tian2010,Okada2006,Shergill2002,Skipper2005,VandeVen2009,Venezia2016}. It is thought that SP initiates in the auditory regions for direct processing of ongoing speech and ultimately maps the speech units into an articulatory network via a sensorimotor interface \citep{Hickok2014,Hickok2004}. Speech production, on the other hand, initiates as an articulatory motor expression, which, through the same sensorimotor interface, becomes transformed into auditory sensory targets in the temporal lobe \citep{Hickok2014,Tian2012}. Although the directionality between the two tasks may be opposed, they have been consistently shown to draw activations from common brain areas. Namely, they seem to converge and produce similar activation patterns largely in phonological networks where the fundamental contrastive speech units (phonemes) are realized \citep{Tian2010,Hickok2011,Hickok2004,Hickok2009,Okada2006,Okada2018}. These studies critically highlight that CS and SP recruit activity from common brain regions, which likely subserve common functions across tasks. 

These source localization studies invite the question, do CS and SP utilize frequency bands in a similar manner? In the brain, information transmission is characterized at various temporal and spatial scales through neural oscillations in the $\delta$-(1-2.5Hz), $\theta$-(4-7Hz), $\alpha$-(8-11Hz), $\beta$-(13-30Hz), and $\gamma$-(30-60Hz) bands \citep{Gross2013,Luo2007,Giraud2007,DiLiberto2015,Giraud2012,Poeppel2020}. In SP, it has been established that lower frequency activity (e.g. $\theta$) detects syllabic quantities through tracking of the speech envelope, whereas higher frequency (e.g. $\gamma$) parses temporally fine units of speech such as phonemes. In addition, the fluctuating $\theta$ phase has been found to modulate the amplitude bursts of the $\gamma$-band by imbuing a rhythmicity to the signal through phase-amplitude coupling (PAC); which enables the coordinated sampling of syllabic and phonological speech items \citep{Hyafil2015,Assaneo2018,Hermes2014}. In speech production, active sensing - a theorized predictive processing mechanism via a motor sampling of sensory faculties - would suggest that overt speech and its variants use neural oscillations similarly to SP since acts of speech production effectively produce self-generated speech noises \citep{Morillon2015}. For instance, differential $\gamma$-band augmentations have been observed in phonological processing regions during overt and covert phoneme repetition tasks, suggesting a possible common role of $\gamma$ activity to SP \citep{Fukuda2010,Toyoda2014}. More generally, dorsal stream motor areas have been found to have its own preferred rhythm of speech production \citep{Restle2012,Assaneo2018,Poeppel2020}, suggesting that the quasi-rhythmicity of the vocal tract articulators generates the cadence of the speech envelope which, in turn, improves speech intelligibility during comprehension \citep{Giraud2000,Boemio2005,Trouvain2007}. Hence, it is increasingly possible that neural oscillations in SP and CS serve similar functions.



However, no studies to date have directly investigated the relative oscillatory contributions and differences in CS and SP. The additive oscillatory components together form the broader characteristics of the signal, and thus efforts to produce models of CS based on signals from SP must first delineate the oscillatory differences across tasks. One such method of understanding how frequency characteristics differ across classes is through a $\mathtt{t}$-test of complex-valued coefficients of a time-frequency transform. This method is referred to as a studentized continuous wavelet transform ($\mathtt{t}$-CWT) and was first introduced by \citeauthor{Bostanov2004} (\citeyear{Bostanov2004}) with the intention of improving feature extraction methods for classification in BCI paradigms. Since then, it has been used in the classification of motor imagery \citep{Darvishi2007,Hsu2007} and the analysis of ERPs in real and simulated EEG data \citep{Real2014}. According to the latter study, distinguishing between ERPs with $\mathtt{t}$-CWT produced high specificity and sensitivity under various signal-to-noise ratios in comparison to common peak detection methods. Studentizing time-frequency information in this manner allows for a direct statistical comparison of time-frequency information between two classes in order to detect frequency indices which are significantly different. Furthermore, compared to the discrete wavelet transform or fast Fourier transform, CWT is an ideal candidate for time series analysis due to its more fine-grained resolution and temporal stability with respect to frequency \citep{Kimata2018}. It is therefore sensible to implement $\mathtt{t}$-CWT with a recording modality with strong temporal resolution such as EEG. 



In the present study, we asked whether $\mathtt{t}$-CWT can identify the frequency bands which are used to distinguish words within CS and SP, comparatively. Considering that CS lacks overt vocalization and thus salient self-stimulation, we hypothesized that lower frequency elements would be overshadowed by high frequency activity, such as $\beta$ and $\gamma$. Subsequently, we asked whether the more pertinent oscillations of CS perform similar functions to those of SP, namely by testing for $\theta$-$\gamma$ PAC. However, in the very likely scenario that the $\theta$-band does not play a major linguistically-relevant role in CS, we tested whether CS's $\gamma$ activity is 'pseudo-coordinated' (or pseudo-coupled) to SP's $\theta$ activity. Such a coupling would indicate that CS's $\gamma$-band response has a rhythmicity that is related to the putative tracking of syllabic quantities by SP's $\theta$ activity, thereby asserting that the $\gamma$-band performs a similar function across tasks. Therefore, we hypothesized that the roles of oscillations in SP and CS would be similar and different in the $\gamma$- and $\theta$-bands, respectively. The remainder of the paper is organized as follows: in Section 2, we provide an overview of the previous works on neural oscillation in CS and SP. In Sections 3 and 4, we provide the details of the study methodology and describe the results. In section 5 we discuss the findings of the study and finally conclude our paper in section 6.

\section{Neural oscillations in speech perception and covert speech}
Over the past two decades, neural oscillations have proven to be a window to understanding a wide variety of cognitive processes. The most important function of neural oscillations is to allow the brain to operate at multiple temporal and spatial scales such that information can be integrated into a holistic percept \citep{Buzsaki2004,Buzsaki2004a,Morillon2015}. Indeed, oscillations are thought to provide the most energy-efficient physical mechanism for synchrony and temporal coordination \citep{Mirollo1990}. In the study of language processing, the induction of such oscillations (e.g. $\delta$, $\theta$, $\alpha$, $\beta$, $\gamma$) contributes to synchronized activities across spatially segregated neuronal assemblies for the coordinated processing of speech units at varying scales \citep{Bastiaansen2006,Weiss2003}.

A multitude of studies have investigated the role of oscillations during SP, and a commonly synthesized interpretation from literature is that SP \textit{multiplexes} neuronal oscillations; that is, SP dynamically samples incoming acoustic information at multiple time scales simultaneously \citep{Gross2013,Luo2007,Ding2017,Giraud2007,Poeppel2020}. A general rule of thumb is that the higher the frequency of oscillation, the finer the detail to which speech information is sampled. For instance, $\delta$ oscillations (1-2.5Hz) have been implicated in the processing of words, phrases, and sentences \citep{Giraud2007,Morillon2012,Doelling2014,Ding2015}. $\theta$ activity (4-7Hz) has been found to be critically sensitive to syllabic modulations namely by tracking the ongoing speech envelope that contains a 4.5Hz syllabic speech rate \citep{Luo2007,Giraud2007,Ghitza2013,Doelling2014}. In contrast, high frequency $\gamma$ activity (30-60Hz) has been found to index processing at the phonemic level, as intracortical studies have located a 'phonotopic' map of phonemes in regions of the superior temporal gyrus producing differential $\gamma$-band augmentations to phonemes  \citep{Chang2010,Moses2016,Pasley2012}. Although $\beta$ activity (13-30Hz) has been commonly associated with motor-related potentials in motor imagery studies, in language, this oscillation is thought to be involved in playing a simultaneous role alongside $\theta$ and $\gamma$ activity by conjoining phonological units into a broader syllabary by binding the activity of temporally segregated neuronal assemblies \citep{Bastiaansen2010,Weiss2003,Weiss2012}. 

Of these oscillations, the most crucial oscillations seem to be the $\theta$ and $\gamma$ band. Evidence for this comes from numerous observations that the phase and amplitude of these frequency bands are coupled in order to synchronize the detection of syllabic boundaries and the parsing  of phonemes \citep{Lizarazu2019,Mai2016,Gross2013,Morillon2012}. In other words, $\theta$ activity samples the input spike trains (induced by speech waveform) to generate basic units and time references of speech for subsequent, finer-detailed processing by $\gamma$ activity \citep{Giraud2012}. The purpose behind this phase-amplitude coupling (PAC) may be to temporally localize $\gamma$ processing power to more descriptive parts of syllabic sound patterns that constitute reference time frames \citep{Hyafil2015}.

Thus, SP multiplexes in relevant frequency bands in order to detect incoming speech and parse them for necessary speech comprehension \citep{Pickering2013}. Although the main oscillatory contributions during SP have been fleshed out, there are vastly fewer studies investigating the role of oscillations in CS. The main feature that distinguishes CS from SP is corollary discharge. Corollary discharge is, in essence, a neural sensory prediction of the consequences of self-generated movements \citep{Wolpert2000,Cullen2004} and, in the case of speech, it is regarded as an auditory prediction of self-generated speech noises \citep{Ford2005,Ford2019,Jack2019,Scott2013}. The sequential estimation mechanism by \citeauthor{Tian2012} (\citeyear{Tian2012}) theorizes that the principal reason why CS produces similar activation patterns as SP in temporal regions of the brain \citep{Tian2010} is due to an auditory prediction of imagined articulation. Indeed, this corollary discharge during CS has been shown to be temporally precise and content-specific \citep{Jack2019}, sensory in nature \citep{Scott2012,Scott2013}, and cancel out self-generated sounds \citep{Okada2018}. Therefore, CS and SP are bridged by a common sensory goal in the auditory domain. This fact invites the question: does CS use oscillations in a similar manner to SP?

The corollary discharge during speech production has been linked to a fronto-temporal $\gamma$-band synchrony \citep{Chen2011}. Moreover, investigations into auditory verbal hallucinations (AVH) have revealed that schizophrenic individuals exhibit significantly suppressed fronto-temporal $\gamma$ synchrony, suggesting that an aberrant corollary discharge is responsible for thoughts manifesting as phantom perceptions \citep{Uhlhaas2006,Uhlhaas2010,Gallinat2004a,Ford2005,Ford2019,Mathalon2008}. More relevant to neurolinguistics, intracranial recording studies of overt and covert phoneme repetition tasks have observed differential $\gamma$ band augmentations in temporal brain regions thought to be responsible for phonological processing \citep{Fukuda2010,Toyoda2014}. As the purpose of corollary discharge is to match the sensory consequences of self-generated actions, these results invite the hypothesis that the corollary discharge produced during CS, reflected in its $\gamma$-band response, may be phonological in nature similar to SP. 

Although these studies provide indirect evidence for a common $\gamma$-band function across tasks, the same may not be the case for $\theta$. \citeauthor{Hermes2014} (\citeyear{Hermes2014}) showed that $\theta$-$\gamma$ PAC is suppressed during CS, with $\theta$ power being anti-correlated to high frequency activity. In contrast to findings of increase PAC in SP studies, the authors surmised that when there is no external input, brain areas may need to downregulate $\theta$ activity in order to allow local neuronal processing \citep{Schroeder2009}. Therefore, it is increasingly possible that the $\theta$-band may play an alternative role to syllabic chunking seen in SP.

EEG is an appropriate modality in which to measure and analyze such brain oscillations due to its fast temporal resolution and ease of setup. In speech processing studies, this modality has been frequently used characterize oscillations in phase entrainment \citep{Zoefel2016}, processing asymmetry \citep{Morillon2012}, speech intelligibility \citep{Onojima2017}, semantic evaluation of speech \citep{Shahin2009}, categorical processing \citep{Bidelman2015a}, and finally oscillatory abnormalities in schizophrenic individuals \citep{Uhlhaas2010,Ford2005}. Although the spatial resolution for EEG is poor due to volume conduction effects, its strong temporal resolution presents this modality as an optimal medium for tracking fast temporal dynamics of ongoing neural oscillations. 

In summary, the findings outlined above demonstrate that certain oscillations may be functionally correlated or divergent in SP and CS. However, no studies have yet determined the relative oscillatory engagements across the two tasks. Hence, the present study used EEG to demonstrate how CS utilizes oscillations relative to SP and whether the most pertinent frequency bands (i.e. $\theta$, $\gamma$) perform similar functions across tasks.

\section{Methods}

\subsection{Participants}
Ten adults between the ages of 20 and 40 without disabilities or known health conditions were recruited for this study (7 Female, age X +/-Y; 3 Male, age X+/-Y). All participants were right-handed to ensure a consistency in the hemispheric dominance of neurolinguistic processing. Furthermore, all participants were native English speakers (i.e. first language). The research ethics board of the Bloorview Research Institute approved this study. Participants provided informed written consent forms.

\begin{figure*}
\centering
\includegraphics[scale=.30]{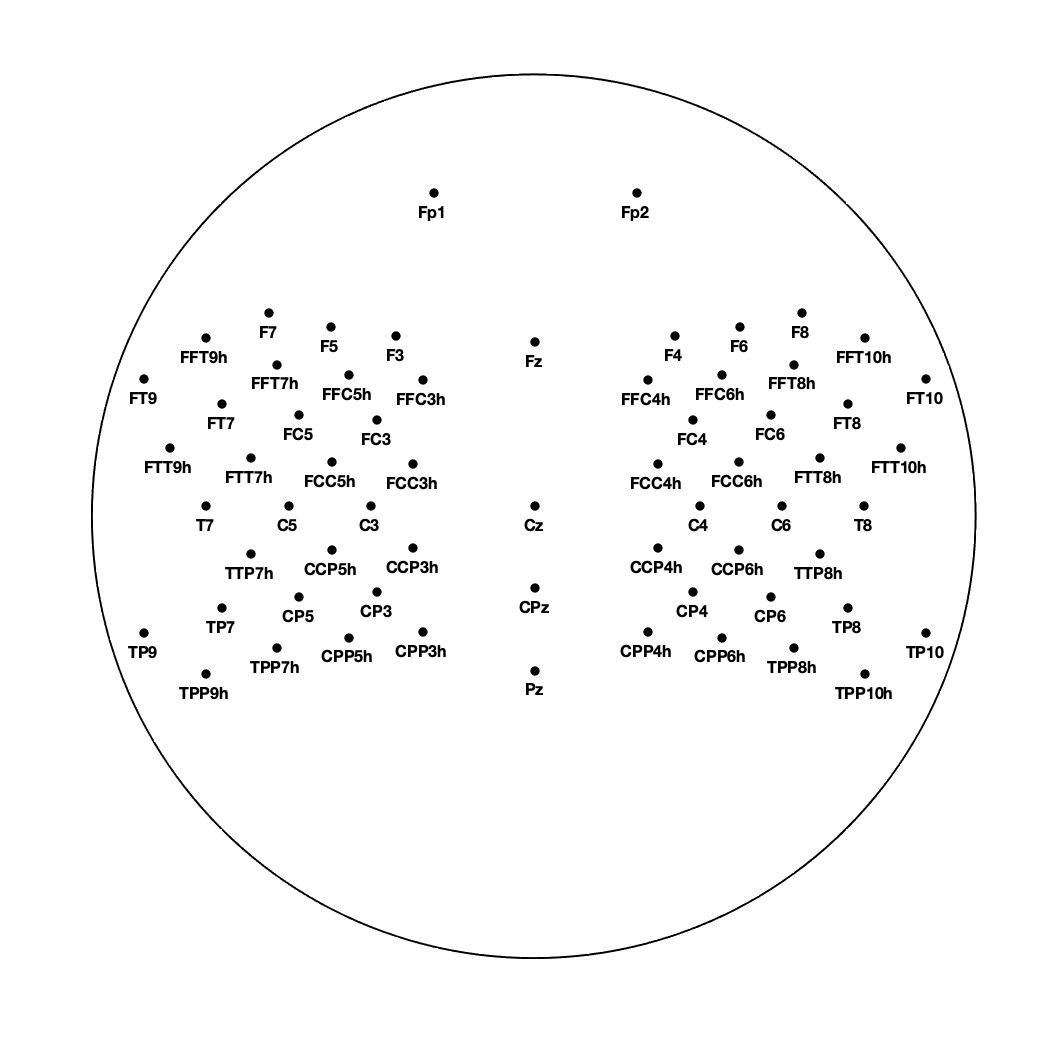}
\caption{64 channel ActiCap wet EEG montage.}
\label{fig:montage}
\end{figure*}

Participants donned a 128 electrode ActiCap EEG cap. Of these 128 channels, 64 were utilized (Fig. \ref{fig:montage}), with the ground electrode at AFz and the reference electrode at FCz. Channels Fp1 and Fp2 were used as ocular artifact detectors. Electrode coverage included the frontal, temporal, and temporo-parietal areas on both hemispheres, including midline components such as Fz, Cz, CPz, and Pz). Data was sampled at 1000Hz and collected through BrainVision Recorder.


\subsection{Experimental procedure}

\begin{figure}
\centering
\includegraphics[scale=.55]{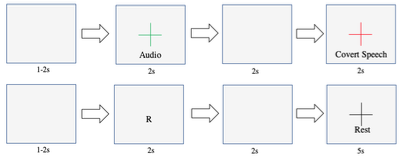}
\caption{The experimental protocol for SP, CS, and rest. SP and rest were preceded with a jitter of 1-2 seconds and followed by either the audio or the symbol, R. In the former case, this was succeeded by a 2 second blank screen and a 2 second window for CS, demarcated by the red cross. In the latter case, the 2 second blank screen was followed by a black cross signalling rest for 5 seconds.}
\label{fig:protocol}
\end{figure}

Each participant was seated comfortably approximately 50cm from the computer screen with a refresh rate of 75Hz. The screen was positioned in the central field of vision and light in the data collection room was turned off prior to beginning the computer task to minimize peripheral vision distractions. Prior to the experiment, prompted by a constant green cross, baseline signals (i.e. neural activity at rest) were recorded for a minute. During the session, SP and CS trial pairs were presented sequentially (Fig. \ref{fig:protocol}). First, a blank screen with a duration jitter between 1-2 seconds was presented, followed by a green cross for 2 seconds. During this time, the audio of the speech token (‘Blue’ or ‘Orange’) was presented once. Succeeding this was another blank screen for 2 seconds, followed by a red cross for 2 seconds. Participants were instructed to covertly rehearse the speech token that they had just heard. Therefore, every SP trial was succeeded by a CS trial with the same word. For the rest trial, a letter R cued the participants to refrain from the speech task and fixate on a black cross held on the screen for 5 seconds. Each block consisted of 10 SP-CS trial pairs and 10 rest trials, and each session consisted of 5 blocks. Therefore, each session involved 50 trials/class, and participants underwent two sessions each at least 2 days apart and at roughly the same times of the day. Each session took approximately one hour.

  
\subsection{Speech items}
The two speech items (‘Blue’, ‘Orange’) were chosen as they differ in the number of syllables and phonemes and differ in their place and manner of articulation. As such, they were expected to engender substantially different patterns of neural activity that could be suitable for detecting substantial differences in the neural activity associated with each item. The rationale was that these variations would encourage the activation of different motor, somatosensory, and auditory neural representations, thereby enhancing signal discriminability. Speech stimuli were generated by Google Cloud Text-to-Speec platform and presented at an approximate rate of 150 words per minute, which is within the range of the natural speaking rate \citep{Giraud2007,Luo2007}. Phoneme models were generated through the Montreal Forced Aligner \citep{McAuliffe2017}. 
 
\subsection{Preprocessing}
Raw data were analyzed in EEGLAB \citep{Delorme2004}. A 4th order zero-phase high pass Butterworth filter with a cutoff frequency of 1Hz was applied to remove baseline drift. Subsequently, a 4th order zero-phase low pass Butterworth filter with a cutoff frequency of 60Hz was applied to remove high frequency noise, as well as the high-$\gamma$ band. Subsequently, the PREP pipeline \citep{Bigdely-shamlo2015} was applied to remove line noise, detect noisy or outlier channels, and to interpolate bad channels. Eye movement artifacts and muscular artifacts were removed in two separate steps using blind source separation through the EEGLAB plugin "Automatic Artifact Removal toolbox". Following preprocessing, a spline Laplacian was applied to establish local relationships between surface potentials and the underlying source activity \citep{Babiloni2001}. Data were downsampled to 256Hz prior to epoching. Data from sessions 1 and 2 were combined. There were a total of 5 classes: SP Blue (SPB), SP Orange (SPO), CS Blue (CSB), CS Orange (CSO), and rest (RST).
 
\subsection{EEG signal processing}

To determine how oscillations drive the distinction of words in CS and SP, we employed a $\mathtt{t}$-CWT routine in order to find the frequency and time indices at which two sets of wavelet coefficients were significantly different. First, each epoched trial was zero-padded with 12 samples in the begining and end. A CWT was conducted on each signal yielding a 55 frequency by 536 time sample matrix (frequencies above 60Hz were removed). CWT was performed using equation (1):

\begin{equation}
W(s,t)=\frac{1}{\sqrt{s}} \int_{-\infty}^{\infty}f(\tau) \psi(\frac{\tau-t}{s})  d\tau
\end{equation}

Where \textit{W(s,t)} represents the wavelet coefficients, \textit{s} denotes the scale or frequency, \textit{t} denotes the time shift, and $\psi$ is the wavelet function which has a zero mean. CWT is thus a sort of template matching computation whereby the cross-covariance between the signal and mother wavelet (here, a Morlet wavelet) is measured by shifting back and forth the latter at dilated and constricted scales. The local extrema of \textit{W(s,t)} signify the points in frequency and time that are best matched between the signal and template wavelet, and can be visualized in the form of a time-frequency plot, referred to as a scalogram.

To determine which mother wavelet suited the data best, each mother wavelet (Bump, Haar, Morse, Morlet) was used to create wavelet coefficients which were then used to reconstruct the signal via inverse CWT. The correlation between the original and reconstructed signals was calculated by conducting cross-correlation tests, divided by the auto-correlation of the original signal. This analysis yielded a metric that showed how similar the two sets of waveforms are related. It was found that the Morlet wavelet correlated best compared to all other mother wavelets. This is consistent with reports that this wavelet is useful for the detection of salient oscillations \citep{Ende1998,Senkowski2002}.

\begin{figure*}[!htbp]
\centering
    \includegraphics[width=0.6\linewidth]{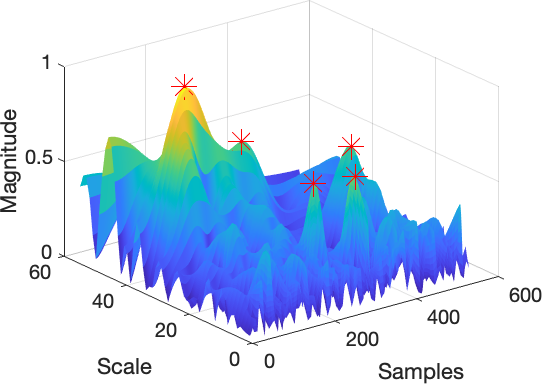}
    \caption{Magnitude scalogram of a $\mathtt{t}$-statistic. Red stars denote the time and scale indices of the peaks of the magnitude scalogram. Peaks were chosen under the criteria that they would exist inside the cone of influence, and importantly, where H=1 (i.e. significant difference).  }
    \label{fig:tstat}
\end{figure*}

For each classification type and the 58 chosen channels, aggregated two-sample $\mathtt{t}$-tests were conducted on the wavelet coefficients across all trials, yielding 55 frequency x 536 sample $\mathtt{t}$-statistic and H (hypothesis test; 0 or 1) matrices. For each channel, the $\mathtt{t}$-CWT was calculated by:

\begin{equation}
    \mathtt{t}^k(s,t) = \frac{\overline{W_x^k(s,t)}-\overline{W_y^k(s,t)}}{\sqrt{\frac{\sigma_x^2}{n} +\frac{\sigma_y^2}{m}}}
\end{equation}

where \textit{k} is channel and $\overline{W_x^k(s,t)}$, $\overline{W_y^k(s,t)}$, $\sigma_x^2(s,t)$, and $\sigma_y^2(s,t)$ denote sample means and standard deviations of wavelet coefficients across trials at each scale \textit{s} and time \textit{t}, with sample sizes \textit{n} and \textit{m}.  Studentizing the wavelet coefficients in this manner enabled the statistical comparison of two classes, describing the frequency and time indices at which they are significantly different. As such, the complex-valued $\mathtt{t}$-statistic matrices served as time-frequency scalograms with greater magnitudes denoting greater differences. Thus, the absolute values of the $\mathtt{t}$-statistic matrices were calculated and subsequently normalized to determine the regional maxima, with the condition that the maxima must be located within the cone of influence, but importantly, where H=1; i.e. where there is a significant difference between classes (Fig. \ref{fig:tstat}). Detecting maxima only within the cone of influence mitigated the risk of detecting artifactual maxima in the scalograms. Each channel produced a different amount of regional maxima. 

\subsection{Feature extraction and classification}

Wavelet coefficients of the CWT output were extracted at the frequency and time indices obtained through t-CWT. These values were extracted for each of the 58 channels and appended into a complex feature matrix. Subsequently, the real and imaginary values of these complex wavelet coefficients were obtained and appended side by side to form a matrix of trials x features, with the last column representing the class identity. For each participant, the original feature matrix was 100 trials x ~500 features approximately. However, Minimally Redundant Maximally Relevant (mRmR) feature selection was conducted to select the top 20 features, which was the approximate cutoff point for feature importance during mRmR. Subsequently, a 10-fold cross validation was conducted on the 100 trials x 20 features matrices and subsequently classified through a support vector machine (SVM) with a radial basis function kernel. All classification types were binary.

\subsection{Performance evaluation}

Average classification accuracies and their standard deviations were obtained using a 10 fold cross-validation. Precision, Recall, and F1-score were calculated for each of these folds and subsequently averaged.

\begin{equation}
    Precision = \frac{True Positives}{True Positives + False Positives}
\end{equation}

\begin{equation}
    Recall = \frac{True Positives}{True Positives + False Negatives}
\end{equation}

\begin{equation}
    F1-score = 2\cdot\frac{Precision * Recall}{Precision+Recall}
\end{equation}

\subsection {Measuring phase-amplitude coupling (PAC)}

The importance of PAC is well documented in speech processing studies \citep{Giraud2012,Hyafil2015,Assaneo2018,Voytek2013,Hermes2014}. To determine significant $\theta$-$\gamma$ coupling in CS, the event-related phase-amplitude coupling (ERPAC) toolbox was utilized \citep{Voytek2013}. A 4th-order Butterworth filter was applied between 4-7Hz to obtain $\theta$-band signals, after which the angle of the Hilbert transform was taken to obtain phase data. The $\gamma$ band was calculated by filtering the signals between 30-60Hz, with a Butterworth filter order of 3*r where r is the sampling rate divided by the low frequency cutoff of the filter, rounded. The $\gamma$-band amplitude was obtained by taking the absolute value of the Hilbert transform. In a specific channel, PAC for each phase-amplitude pair was calculated through the \textit{circ\_corrcl.m} function from the CircStat toolbox \citep{Berens2009} at each timepoint and across all trials. This function calculates the correlation coefficient ($\rho$) between a circular/angular ($\phi,c,s$) and linear random variable (such as amplitude - $a$) by linearizing the phase variable into sin and cosine components:

\begin{equation}
    \rho_{\phi a} = \sqrt{\frac{r^2_{ca} +r^2_{sa} -2r_{ca} r_{sa} r_{cs}}{1-r^2_{cs}}}
\end{equation}

where $r_{ca} = corr(cos\phi[t],a[t])$, $r_{sa} = corr(sin\phi[t], a[n])$, $r_{cs} = corr(sin\phi[t], cos\phi[t])$, and $corr(x,y)$ is the Pearson correlation between $x$ and $y$ with the assumption that the distribution of $x$ and $y$ are Gaussian. $\phi[t]$ and $a[t]$ are the instantaneous phase and instantaneous analytic amplitude, respectively. Utilizing this function enabled the assessment of relationships between circular $\theta$ phase and linear $\gamma$ amplitude at each time point and across trials.

 Subsequently, 1000 surrogate runs were conducted by shifting the trials of the amplitude data and testing correlation between the phase data and shifted amplitude data across trials at each time point. These PAC $\rho$ values were compared by first applying Fisher's \textit{z}-transform to normalize the correlation coefficients:
 
 \begin{equation}
     z_{rt}=\frac{1}{2} ln\left(\frac{1+\rho_t}{1-\rho_t}\right)
 \end{equation}
 
 and calculating the difference between \textit{z}-transformed coefficients:
 \begin{equation}
     \triangle\rho_z = z(\rho_{true}) - z(\rho_{surrogate})
 \end{equation}
 
 From this, the \textit{z}-score can be calculated by:
  \begin{equation}
      z=\frac{\triangle\rho_z}{\sigma}
  \end{equation}
 
where $\sigma$ is standard error. \textit{z}-scores were then transformed into \textit{p}-values via a normal cumulative distribution function with $\mu=0$, $\sigma=1$. The reported \textit{p}-values denote the time points at which there are significant $\theta$-$\gamma$ PAC occurring against a surrogate population within a specific channel. For cross-task PAC (e.g. SP $\theta$-CS $\gamma$), the class indices used for amplitude calculation was simply switched to a CS class.

 \subsection{Statistical analysis}
For the t-CWT results, the frequency indices were tabulated across all channels and participants and subsequently categorized into the five major bands. Shapiro-Wilks tests were conducted on each category to confirm non-normality. Subsequently, Wilcoxon Rank Sum tests were conducted for unequal medians. 

To test for significant correlations between the $\theta$ phases of SP and CS classes, the \textit{circ\_corrcc.m} function from the CircStat toolbox was invoked \citep{Berens2009}. This function assesses the correlation between two circular/angular random signals:

\begin{equation}
    \rho_{\alpha\beta} =\frac{\sum_i sin(\alpha_i-\overline{\alpha}) sin(\beta_i-\overline{\beta})}{\sqrt{\sum_isin^2(\alpha_i-\overline{\alpha})sin^2(\beta_i-\overline{\beta})}}
\end{equation}

where  $\alpha$ and $\beta$ denote two samples of angular data and $\overline{\alpha}$ and $\overline{\beta}$ denote their means. Under the null hypothesis of no significant correlations, the \textit{p}-value to this correlation was computed by a normally distributed test statistic. This enabled the testing of angular correlation between two sets of SP $\theta$ phase across all trials of participants (1000 trials).




\section{Results}
\subsection{Task-dependent utilization of oscillations for distinction of words}

\begin{table*}[htbp!]

\small
\centering
\scalebox{0.6}{
\begin{tabular}{|c|c|c|c|c|c|c|c|c|c|c|c|}
\hline
\textbf{Part/Type}                &                & \textbf{1} & \textbf{2} & \textbf{3} & \textbf{4} & \textbf{5} & \textbf{6} & \textbf{7} & \textbf{8} & \textbf{9} & \textbf{10} \\ \hline
\multirow{4}{*}{\textbf{SPO-CSO}} & \textbf{Acc.}  & 88.0(8.4)  & 73.5(11.8) & 80.0(8.7)  & 85.5(9.3)  & 89.0(5.4)  & 90.5(7.9)  & 66.5(12.3) & 88.3(7.6)  & 79.5(7.2)  & 87.0(4.0)   \\ \cline{2-12} 
                                  & \textbf{Prec.} & 0.88(0.09) & 0.78(0.13) & 0.82(0.08) & 0.86(0.10) & 0.90(0.06) & 0.91(0.08) & 0.68(0.13) & 0.89(0.08) & 0.81(0.08) & 0.88(0.04)  \\ \cline{2-12} 
                                  & \textbf{Rec.}  & 0.88(0.09) & 0.74(0.12) & 0.80(0.09) & 0.86(0.10) & 0.89(0.06) & 0.91(0.08) & 0.67(0.13) & 0.88(0.08) & 0.79(0.08) & 0.87(0.04)  \\ \cline{2-12} 
                                  & \textbf{F1}    & 0.88(0.09) & 0.76(0.13) & 0.81(0.09) & 0.86(0.10) & 0.89(0.06) & 0.91(0.08) & 0.67(0.13) & 0.89(0.08) & 0.80(0.08) & 0.87(0.04)  \\ \hline
\multirow{4}{*}{\textbf{CSB-CSO}} & \textbf{Acc.}  & 79.5(9.3)  & 71.0(11.8) & 81.5(7.1)  & 82.0(7.8)  & 84.5(5.2)  & 84.0(7.3)  & 69.5(13.7) & 77.2(5.2)  & 81.0(3.0)  & 82.5(7.2)   \\ \cline{2-12} 
                                  & \textbf{Prec.} & 0.80(0.10) & 0.73(0.13) & 0.82(0.07) & 0.83(0.09) & 0.85(0.06) & 0.85(0.08) & 0.70(0.16) & 0.78(0.06) & 0.82(0.04) & 0.83(0.08)  \\ \cline{2-12} 
                                  & \textbf{Rec.}  & 0.80(0.10) & 0.71(0.12) & 0.82(0.07) & 0.82(0.08) & 0.85(0.06) & 0.84(0.08) & 0.70(0.14) & 0.77(0.06) & 0.81(0.03) & 0.83(0.08)  \\ \cline{2-12} 
                                  & \textbf{F1}    & 0.80(0.10) & 0.72(0.13) & 0.82(0.07) & 0.82(0.09) & 0.85(0.06) & 0.84(0.08) & 0.70(0.15) & 0.78(0.06) & 0.82(0.03) & 0.83(0.08)  \\ \hline
\multirow{4}{*}{\textbf{SPB-CSB}} & \textbf{Acc.}  & 86.5(3.9)  & 76.5(12.9) & 88.5(4.5)  & 91.5(5.5)  & 88.0(4.6)  & 94.0(8.9)  & 64.5(13.5) & 84.4(7.4)  & 79.5(8.8)  & 85.5(6.5)   \\ \cline{2-12} 
                                  & \textbf{Prec.} & 0.87(0.04) & 0.78(0.13) & 0.90(0.04) & 0.92(0.06) & 0.88(0.04) & 0.94(0.09) & 0.65(0.17) & 0.86(0.08) & 0.81(0.09) & 0.86(0.07)  \\ \cline{2-12} 
                                  & \textbf{Rec.}  & 0.87(0.04) & 0.77(0.14) & 0.89(0.05) & 0.92(0.06) & 0.88(0.05) & 0.94(0.09) & 0.65(0.14) & 0.84(0.08) & 0.80(0.09) & 0.86(0.07)  \\ \cline{2-12} 
                                  & \textbf{F1}    & 0.87(0.04) & 0.77(0.13) & 0.89(0.04) & 0.92(0.06) & 0.88(0.05) & 0.94(0.09) & 0.65(0.15) & 0.85(0.08) & 0.80(0.09) & 0.86(0.07)  \\ \hline
\multirow{4}{*}{\textbf{SPO-CSO}} & \textbf{Acc.}  & 84.5(8.2)  & 72.5(11.5) & 85.5(6.5)  & 94.5(6.1)  & 90.5(5.7)  & 91.5(6.3)  & 68.5(14.8) & 90.6(7.5)  & 82.5(10.5) & 89.5(4.2)   \\ \cline{2-12} 
                                  & \textbf{Prec.} & 0.86(0.09) & 0.74(0.12) & 0.87(0.06) & 0.95(0.06) & 0.92(0.05) & 0.92(0.06) & 0.70(0.16) & 0.91(0.08) & 0.83(0.11) & 0.90(0.04)  \\ \cline{2-12} 
                                  & \textbf{Rec.}  & 0.85(0.09) & 0.73(0.12) & 0.85(0.07) & 0.95(0.06) & 0.91(0.06) & 0.92(0.07) & 0.68(0.16) & 0.91(0.08) & 0.83(0.11) & 0.90(0.04)  \\ \cline{2-12} 
                                  & \textbf{F1}    & 0.85(0.09) & 0.73(0.12) & 0.86(0.07) & 0.95(0.06) & 0.91(0.05) & 0.92(0.07) & 0.69(0.16) & 0.91(0.08) & 0.83(0.11) & 0.90(0.04)  \\ \hline
\multirow{4}{*}{\textbf{SPB-RST}} & \textbf{Acc.}  & 87.0(7.1)  & 77.0(10.5) & 83.0(7.5)  & 94.5(5.7)  & 82.5(7.5)  & 89.0(5.8)  & 65.5(14.6) & 88.9(9.3)  & 82.5(6.8)  & 87.5(8.7)   \\ \cline{2-12} 
                                  & \textbf{Prec.} & 0.88(0.07) & 0.81(0.09) & 0.84(0.08) & 0.95(0.06) & 0.83(0.07) & 0.90(0.06) & 0.66(0.16) & 0.90(0.10) & 0.83(0.07) & 0.88(0.09)  \\ \cline{2-12} 
                                  & \textbf{Rec.}  & 0.87(0.08) & 0.77(0.11) & 0.83(0.08) & 0.95(0.06) & 0.83(0.08) & 0.89(0.06) & 0.66(0.15) & 0.89(0.10) & 0.83(0.07) & 0.88(0.09)  \\ \cline{2-12} 
                                  & \textbf{F1}    & 0.87(0.07) & 0.79(0.10) & 0.83(0.08) & 0.95(0.06) & 0.83(0.08) & 0.90(0.06) & 0.66(0.16) & 0.89(0.10) & 0.83(0.07) & 0.88(0.09)  \\ \hline
\multirow{4}{*}{\textbf{SPO-RST}} & \textbf{Acc.}  & 90.0(6.3)  & 72.5(10.1) & 82.0(5.6)  & 94.5(4.2)  & 91.5(7.1)  & 92.0(6.8)  & 67.0(12.5) & 91.7(7.6)  & 79.0(7.7)  & 92.5(5.6)   \\ \cline{2-12} 
                                  & \textbf{Prec.} & 0.91(0.06) & 0.74(0.11) & 0.83(0.05) & 0.95(0.04) & 0.92(0.07) & 0.93(0.06) & 0.68(0.13) & 0.92(0.08) & 0.80(0.08) & 0.94(0.05)  \\ \cline{2-12} 
                                  & \textbf{Rec.}  & 0.90(0.07) & 0.73(0.11) & 0.82(0.06) & 0.95(0.04) & 0.92(0.07) & 0.92(0.07) & 0.67(0.13) & 0.92(0.08) & 0.79(0.08) & 0.93(0.06)  \\ \cline{2-12} 
                                  & \textbf{F1}    & 0.91(0.06) & 0.73(0.11) & 0.83(0.06) & 0.95(0.04) & 0.92(0.07) & 0.93(0.06) & 0.68(0.13) & 0.92(0.08) & 0.79(0.08) & 0.93(0.05)  \\ \hline
\multirow{4}{*}{\textbf{CSB-RST}} & \textbf{Acc.}  & 82.5(9.6)  & 72.5(10.5) & 82.5(6.0)  & 81.5(7.4)  & 82.0(6.8)  & 83.5(6.7)  & 71.0(17.4) & 70.6(11.7) & 83.5(7.1)  & 84.0(6.6)   \\ \cline{2-12} 
                                  & \textbf{Prec.} & 0.83(0.10) & 0.76(0.10) & 0.84(0.06) & 0.82(0.08) & 0.83(0.07) & 0.85(0.07) & 0.71(0.20) & 0.72(0.13) & 0.84(0.08) & 0.86(0.07)  \\ \cline{2-12} 
                                  & \textbf{Rec.}  & 0.83(0.10) & 0.73(0.11) & 0.83(0.06) & 0.81(0.08) & 0.82(0.07) & 0.84(0.07) & 0.71(0.18) & 0.71(0.12) & 0.84(0.07) & 0.84(0.07)  \\ \cline{2-12} 
                                  & \textbf{F1}    & 0.83(0.10) & 0.74(0.10) & 0.83(0.06) & 0.82(0.08) & 0.82(0.07) & 0.84(0.07) & 0.71(0.19) & 0.71(0.13) & 0.84(0.08) & 0.85(0.07)  \\ \hline
\multirow{4}{*}{\textbf{CSO-RST}} & \textbf{Acc.}  & 86.5(7.4)  & 65.5(10.1) & 82.0(6.8)  & 81.5(7.1)  & 88.0(9.5)  & 85.0(5.5)  & 63.0(14.2) & 75.0(9.0)  & 81.0(9.7)  & 84.5(6.1)   \\ \cline{2-12} 
                                  & \textbf{Prec.} & 0.87(0.07) & 0.67(0.12) & 0.83(0.08) & 0.83(0.07) & 0.88(0.10) & 0.86(0.05) & 0.66(0.18) & 0.76(0.09) & 0.83(0.11) & 0.86(0.07)  \\ \cline{2-12} 
                                  & \textbf{Rec.}  & 0.87(0.08) & 0.66(0.11) & 0.82(0.07) & 0.82(0.07) & 0.88(0.10) & 0.85(0.06) & 0.63(0.15) & 0.75(0.10) & 0.81(0.10) & 0.85(0.06)  \\ \cline{2-12} 
                                  & \textbf{F1}    & 0.87(0.07) & 0.66(0.11) & 0.82(0.07) & 0.82(0.07) & 0.88(0.10) & 0.86(0.06) & 0.64(0.16) & 0.75(0.09) & 0.82(0.11) & 0.85(0.07)  \\ \hline
\end{tabular}}

\caption{Classification scores for each classification type and for each participant. Post-mRmR feature selection features were classified using a SVM with a radial basis function kernel. Classification accuracies were calculated by averaging across the 10 cross validation folds (standard deviation in brackets). Precision, recall, and F1-scores were calculated for each fold and subsequently averaged.}
\label{tab:accprf}
\end{table*}

 Extracting the real and imaginary values of imaginary coefficients  at these indices produced significantly higher binary SVM classification accuracies than chance level (Table \ref{tab:accprf}) in most participants with the exception of P2 and P7. Similarly, these features were found to have high precision, recall, and F1-score across most participants. The two participants who seemed to perform relatively poorly had greater standard deviations across classifications and lower performance scores.

\begin{figure*}[p]
  \centering
  \begin{subfigure}[b]{0.475\textwidth}  
    \includegraphics[width=\textwidth]{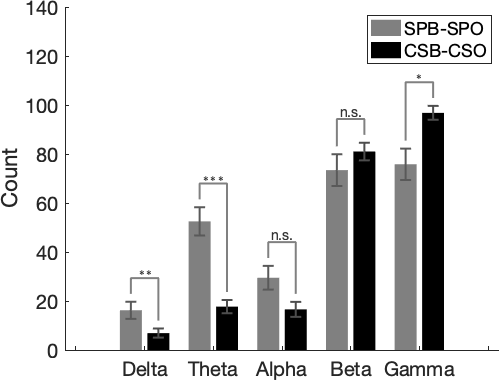}
     \caption{}
  \end{subfigure}
  \begin{subfigure}[b]{0.475\textwidth}  
    \includegraphics[width=\textwidth]{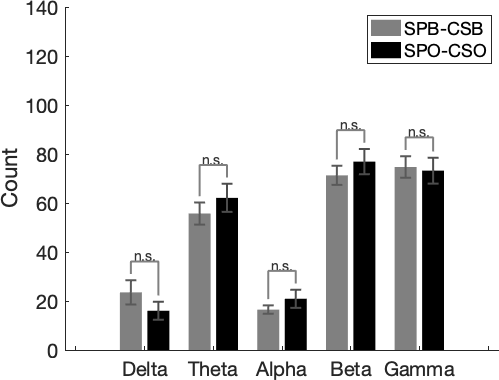}
     \caption{}
  \end{subfigure}
  \begin{subfigure}[b]{0.475\textwidth}  
    \includegraphics[width=\textwidth]{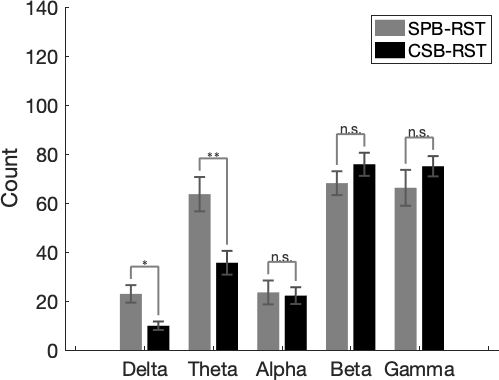}
     \caption{}
  \end{subfigure}
  \begin{subfigure}[b]{0.475\textwidth}  
    \includegraphics[width=\textwidth]{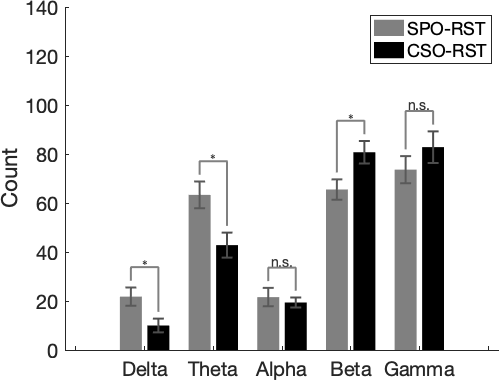}
     \caption{}
  \end{subfigure}
  \caption{Differential utilization of frequency bands in covert speech and speech perception. Each frequency and time index was determined first by conducting a two-sample $\mathtt{t}$-test against two sets of wavelet coefficients. Maxima of the $\mathtt{t}$-statistic were used as indices provided that the hypothesis test returned a statistically significant difference. Data were pooled across all channels and participants. A Shapiro-Wilks test was conducted between each binary classification pair (e.g. SPB vs SPO) to confirm non-normality. Subsequently, a Wilcoxon Rank Sum test was conducted to test for significantly different medians. The distinction of SP classes produced a greater engagement of low frequency $\delta$ and $\theta$, whereas CS involved more low $\gamma$ activity for distinction (a). No significant differences were observed when comparing binary classifications involving corresponding SP and CS classes (b). $\delta$ and $\theta$ activity contributed significantly more to distinguishing SP from rest than CS from rest (c, d). $\beta$ activity contributed significantly more to distinction of CSO vs RST than SPO vs RST (d). (*-\textit{p}<0.05; **-\textit{p}<0.01; ***=\textit{p}<0.001; ns-no significance).}
  \label{fig:freqcount}
\end{figure*}


Assessing the frequency indices at which opposing classes are significantly different (cumulative over all 58 channels and participants) revealed significant differences in oscillatory characteristics between SP and CS (Fig. \ref{fig:freqcount}). Consistent with previous studies \citep{Gross2013,Luo2007}, SP multiplexed in all relevant frequency bands such as $\delta$, $\theta$, $\beta$, and $\gamma$. After confirming non-normality of distribution of frequencies through Shapiro-Wilks tests (\textit{p}<0.05), Wilcoxon  Rank Sum tests revealed that distinguishing SP classes engages significantly more of lower frequency $\delta$ and $\theta$ (\textit{p}<0.01, \textit{p}< 0.0001), whereas distinguishing CS involves more usage of low $\gamma$ (\textit{p}<0.05) (Fig. \ref{fig:freqcount}a). No significant differences in $\alpha$ and $\beta$ were observed (\textit{p}>0.01). Distinguishing between corresponding SP and CS classes (e.g. Blue) showed that SP and CS classes differ similarly across words (\textit{p}>0.05) (Fig. \ref{fig:freqcount}b). The distinction of active classes from rest compared between CS and SP revealed significantly higher involvement of $\delta$ and $\theta$ bands in SP (\textit{p}<0.05, \textit{p}<0.01), whereas CSO vs RST produced significantly higher $\beta$ involvement than SPO vs RST (\textit{p}<0.05) (Fig.\ref{fig:freqcount}c, d).

\begin{figure*}[!htbp]
\begin{subfigure}[b]{0.475\linewidth}
  \centering
  \includegraphics[width=\linewidth]{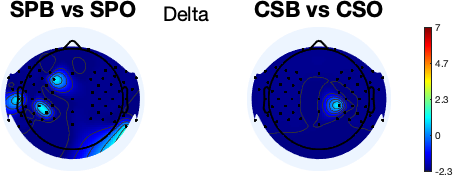}
  \caption{}
  \end{subfigure}
  \begin{subfigure}[b]{0.475\linewidth}
  \centering
    \includegraphics[width=\linewidth]{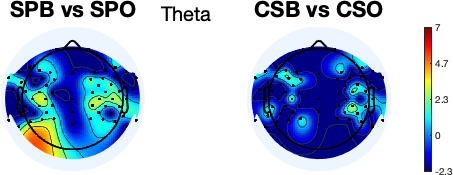}
    \caption{}
  \end{subfigure}
  \begin{subfigure}[b]{0.475\linewidth}
  \centering
  \includegraphics[width=\linewidth]{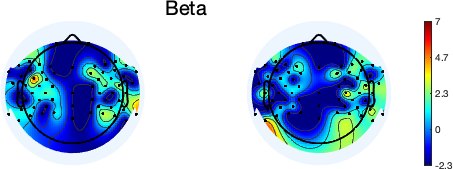}
  \caption{}
  \end{subfigure}
  \begin{subfigure}[b]{0.475\linewidth}
  \centering
    \includegraphics[width=\linewidth]{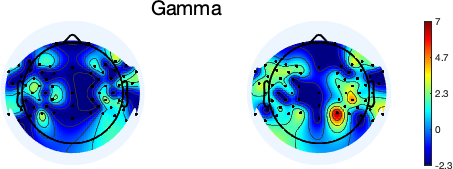}
    \caption{}
  \end{subfigure}
  
 \begin{subfigure}[b]{0.475\linewidth}
  \centering
    \includegraphics[width=\linewidth]{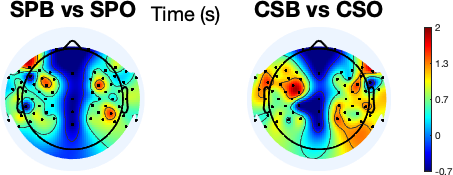}
    \caption{}
  \end{subfigure}
  
  \caption{Topography of frequency indices (a-d) time indices (e) of speech perception and covert speech tasks. (a-d) Calculated from 20 features selected through mRmR feature selection as a tabulated sum across participants. Figures were generated through EEGLAB. \citep{Delorme2004}. Scale shows the number of tabulated distinctions in the frequency band. Brighter colors mean higher count. The distinction of CS classes produces involves higher frequencies whereas the distinction of SP is largely based on lower frequencies. (e) Time indices were calculated by taking the median of time values across participants. Scales show the time in seconds. Brighter and darker colors mean high- and low-latency, respectively. SP and CS both produced low latency distinctions in the temporal and temporo-parietal regions, with high latency activations in the frontal/motor regions.}
  
  \label{fig:fttopo}
\end{figure*}

To determine the topography of oscillatory differences, EEGLAB's \citep{Delorme2004} \textit{topoplot.m} function was used on features selected through mRmR feature selection. Distinctions in channel locations were tabulated (cumulative across participants) for each frequency band (Fig. \ref{fig:fttopo}). SP showed relatively greater inter-participant consistency in the $\theta$-band, with widespread distinctions across temporal and temporo-parietal regions and high counts in this frequency (Fig. \ref{fig:fttopo}b). SP's $\gamma$ distinctions were focally distributed with lower consistencies across participants (Fig. \ref{fig:fttopo}d). In contrast, CS produced focal $\theta$-band distinctions with low count/consistency and widespread $\gamma$-band distinctions with high amount of consistency across participants in temporal and temporo-parietal regions. Both tasks produced comparable amount of $\beta$-band distinctions, but CS produced more counts of distinguishable $\beta$ patterns in the right hemisphere (Fig. \ref{fig:fttopo}c). CS had minimal $\delta$ activity, whereas SP showed three foci of $\delta$ activity in the left hemisphere (Fig. \ref{fig:fttopo}a).


\begin{figure*}[!htbp]
  \centering
  \begin{subfigure}[b]{0.475\linewidth}
    \includegraphics[width=\textwidth]{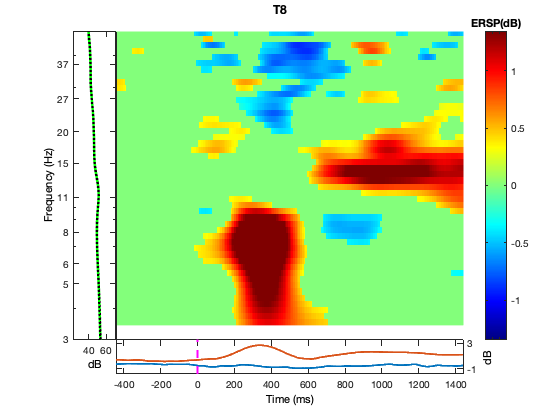}
     \caption{}
  \end{subfigure}
  \begin{subfigure}[b]{0.475\linewidth}
    \includegraphics[width=\textwidth]{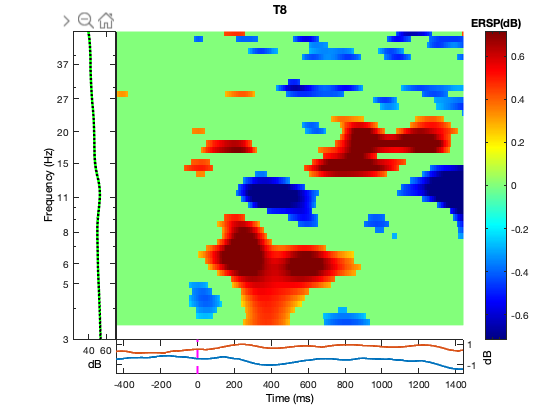}
    \caption{}
  \end{subfigure}
  \begin{subfigure}[b]{0.475\linewidth}
    \includegraphics[width=\textwidth]{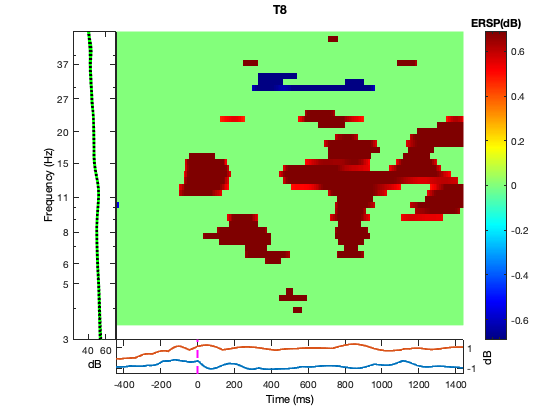}
    \caption{}
  
  \end{subfigure}
  \caption{Event related spectral perturbation in perception (a), covert speech (b), and rest tasks (c) in channel T8. Figures were generated using EEGLAB's \textit{spectopo.m} function \citep{Delorme2004} and data aggregated from participants 1-5. Results are FDR-corrected at \textit{p}<0.05. Frequency units are logarithmically spaced. The ERSP for SP (a) and CS (b) both show early latency synchronization in the $\theta$ band, followed by $\beta$-band synchronization. Rest showed a scattered synchronization of $\alpha$ and $\beta$-bands. }
  \label{fig:ersp}
  \end{figure*}

\subsection{Task-dependent coupling of Theta phase to Gamma amplitude}
  
Event-related spectral perturbations (ERSPs) revealed strong transient synchronization in the $\theta$ band for CS and SP between 200-500ms (Fig. \ref{fig:ersp}a, b). For both tasks, this was succeeded by $\beta$-band synchronization starting at the offset of $\theta$-band synchronization. $\gamma$-band desynchronizations were observed for both tasks between 200-500ms, but more scattered for CS. Rest showed scattered and unorderly synchronizations in the $\alpha$ and $\beta$ bands (Fig. \ref{fig:ersp}c).

\begin{figure*}[p]
  \begin{subfigure}[b]{0.475\linewidth}
    \centering
    \includegraphics[width=\linewidth]{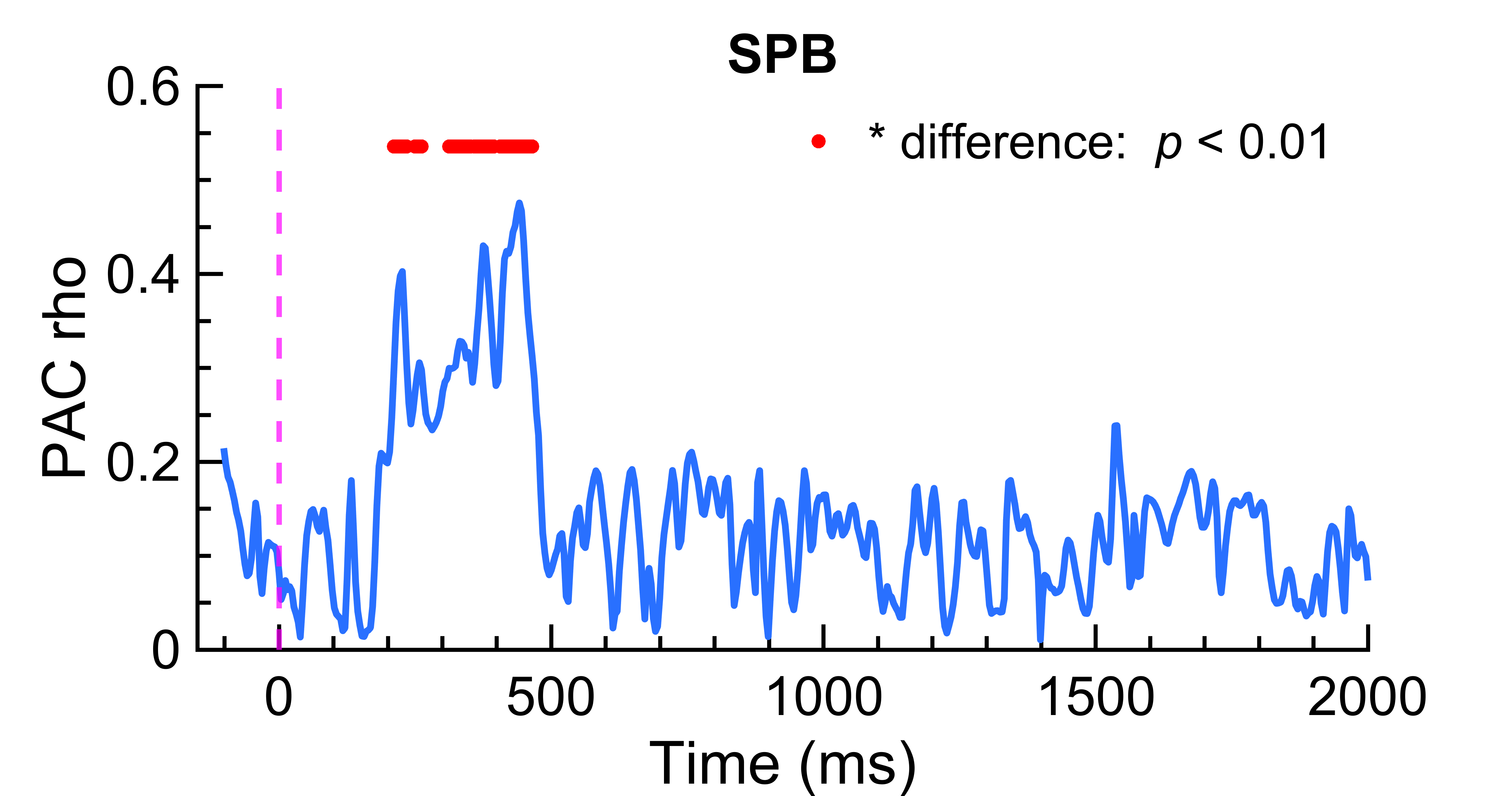}
     \caption{}
  \end{subfigure}
  \begin{subfigure}[b]{0.475\linewidth}
    \centering
      \includegraphics[width=\linewidth]{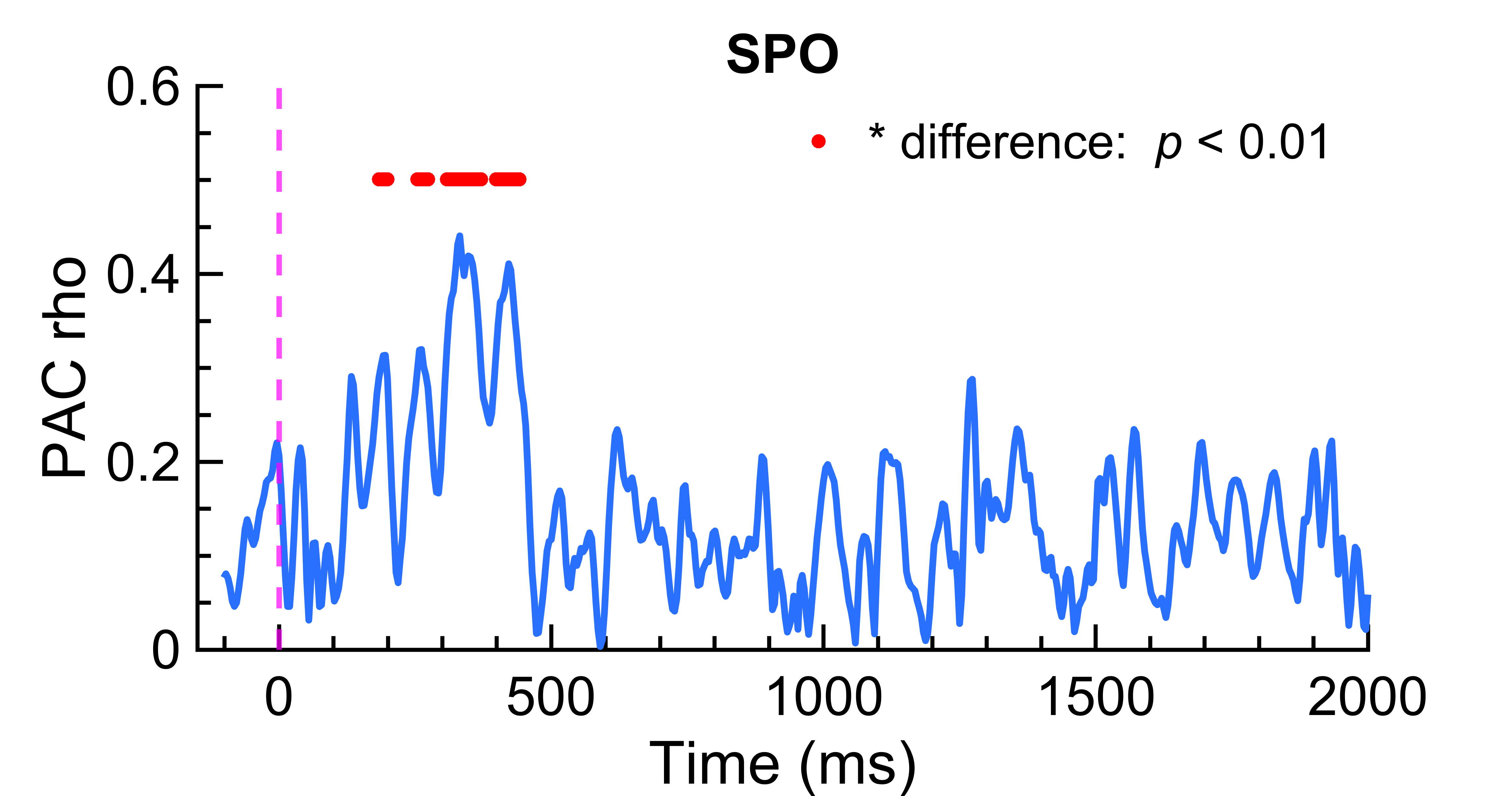}
    \caption{}
  \end{subfigure}
  \begin{subfigure}[b]{0.475\linewidth}
    \centering
    \includegraphics[width=\linewidth]{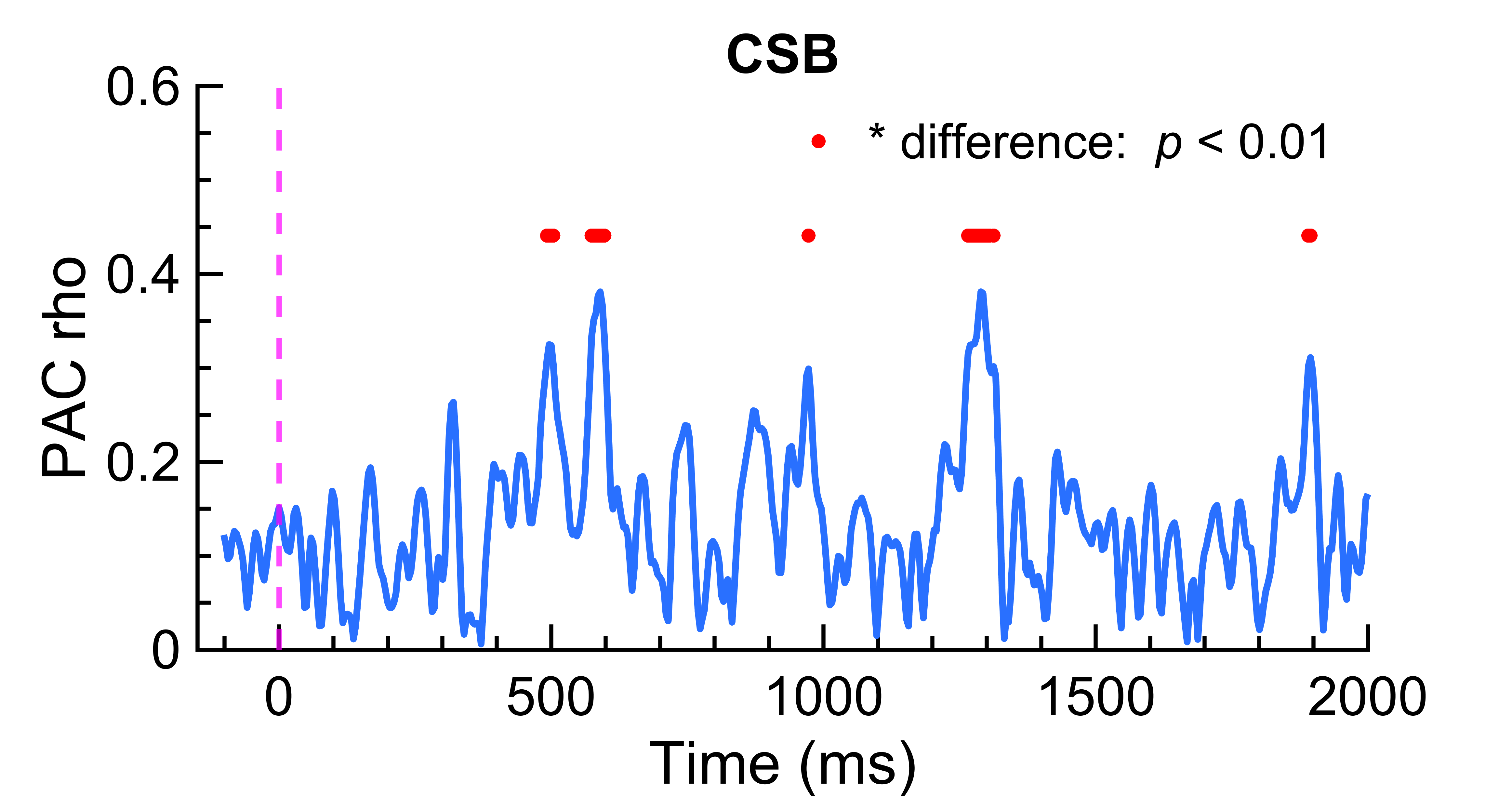}
    \caption{}
  \end{subfigure}
  \begin{subfigure}[b]{0.475\linewidth}
    \centering
    \includegraphics[width=\linewidth]{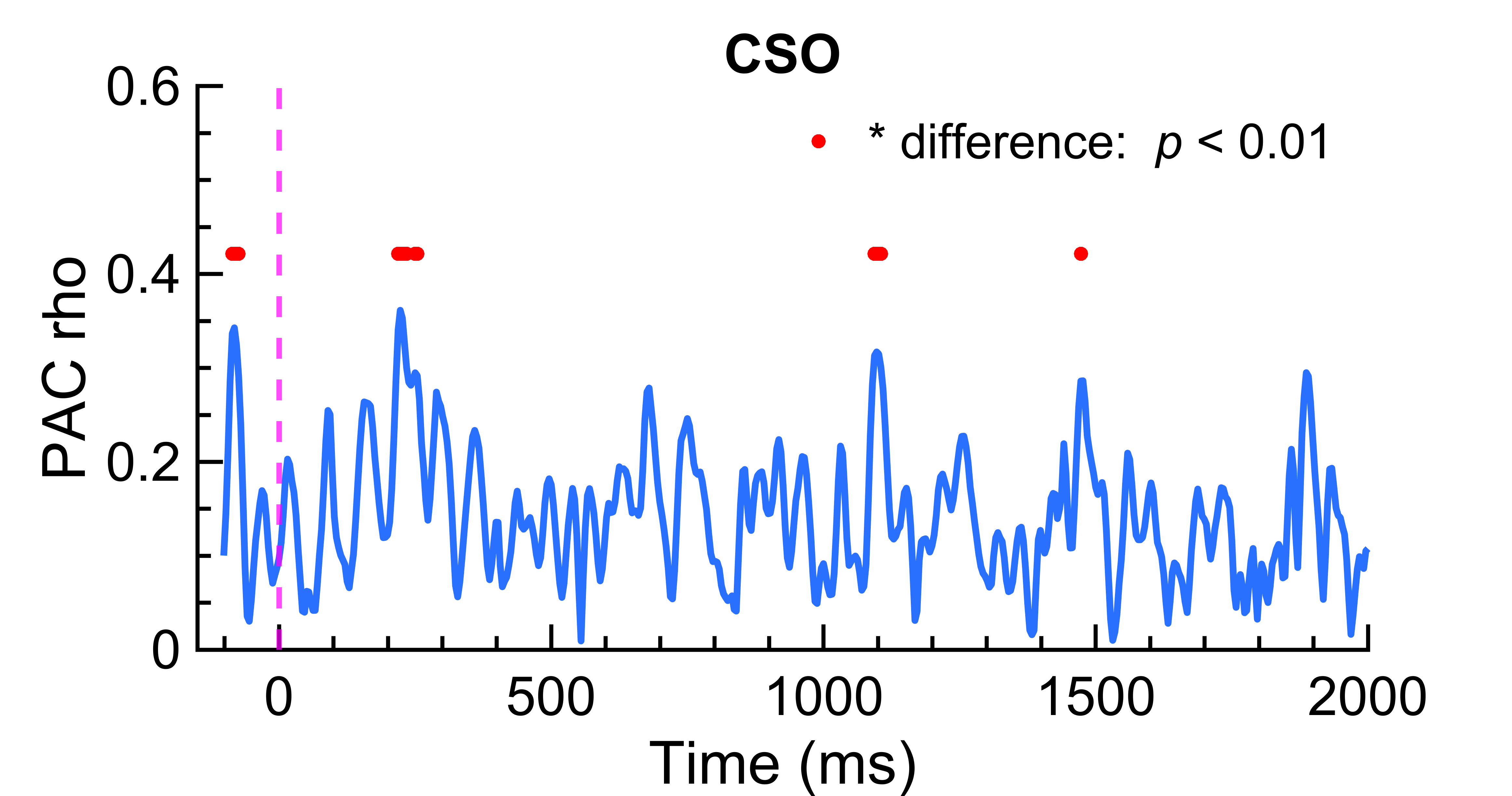}
    \caption{}
  \end{subfigure}
  \begin{subfigure}[b]{0.475\linewidth}
    \centering
    \includegraphics[width=\linewidth]{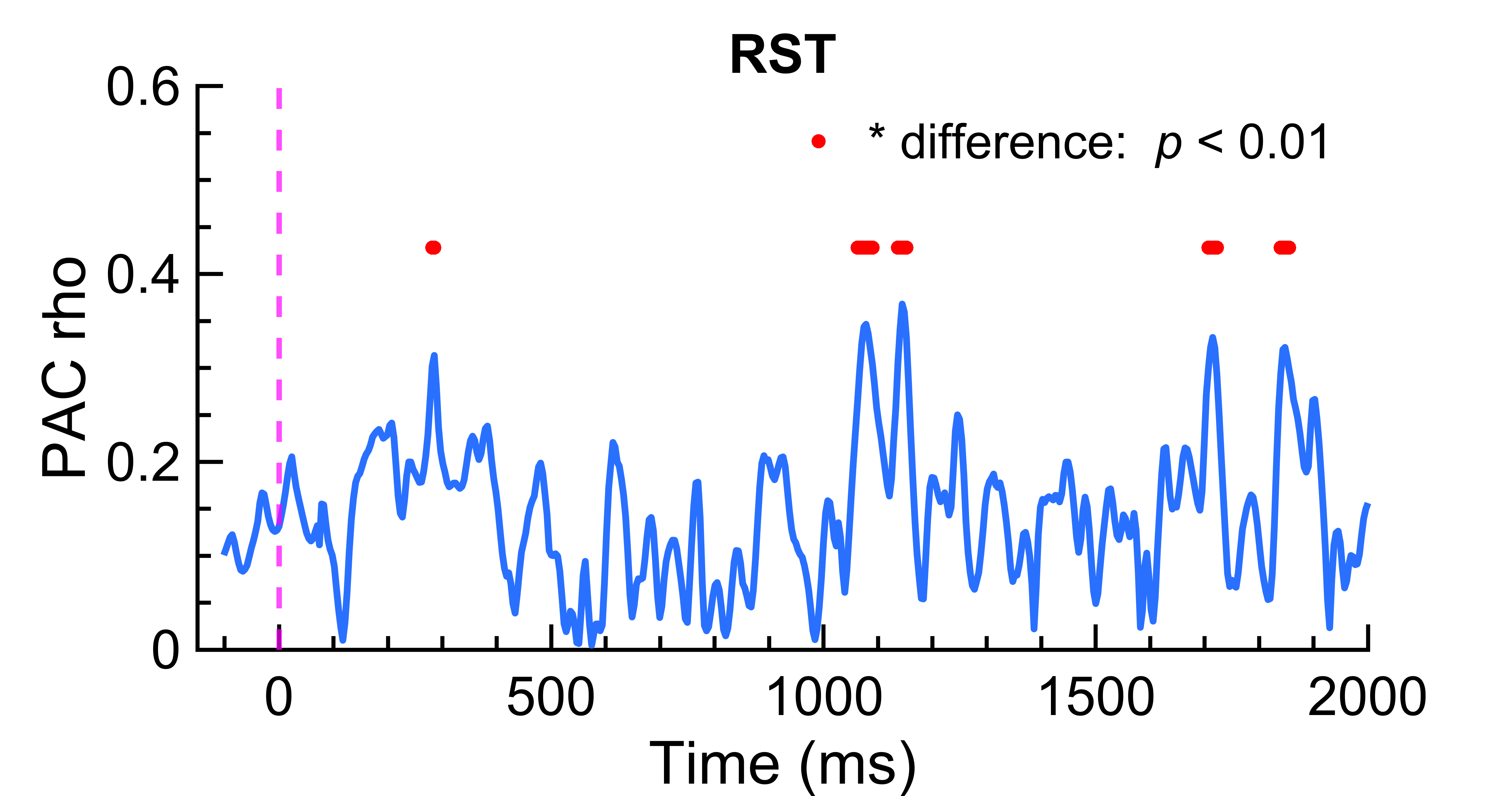}
    \caption{}
  \end{subfigure}
  
  \caption{Speech perception produces significant $\theta$ phase-$\gamma$ amplitude coupling between 200-500ms. Calculated through ERPAC toolbox \citep{Voytek2013}. To determine whether arising PAC relationships index an inter-trial relationship and not an artifact of stimulus-evoked responses, we conducted a resampling analysis (surrogate testing) that randomizes the phase-amplitude relationship across trials. This was done 1000 times per sample and resulted in a distribution of possible surrogate PACs. SP produced significant PAC relationships occurring between 200-500ms, while CS and rest produced relatively little and sparse PAC between $\theta$ phase and low $\gamma$ amplitude. Dotted lines indicate significant difference of true PAC from surrogates (\textit{p} <0.05). Figures were generated from participant 5.}
  
  \label{fig:pac}
  \end{figure*}

\begin{figure*}[htbp!]
  \begin{subfigure}[b]{0.475\linewidth}
  \centering
  \includegraphics[width=\linewidth]{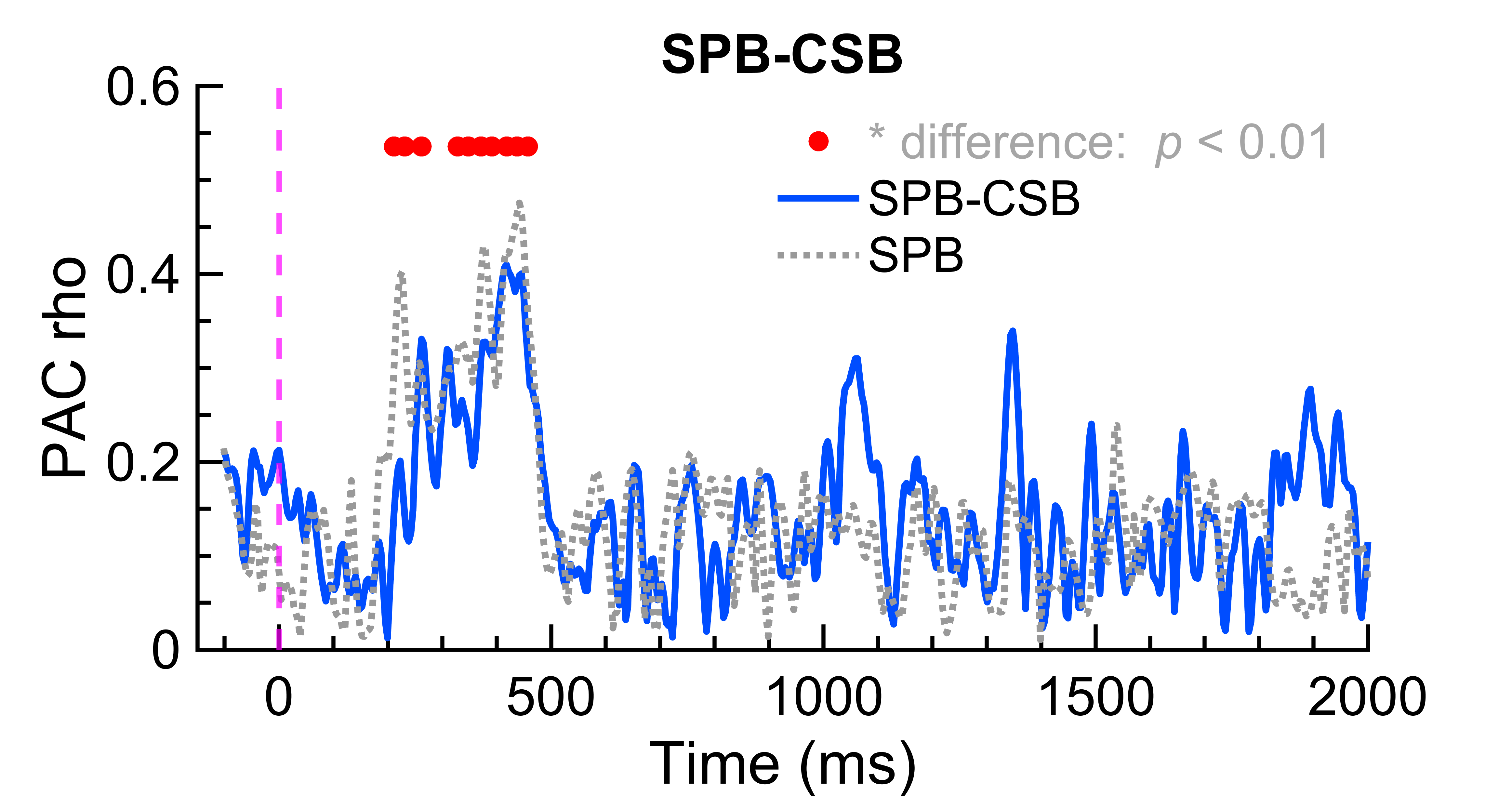}
  \caption{}
  \end{subfigure}
  \begin{subfigure}[b]{0.475\linewidth}
  \centering
  \includegraphics[width=\linewidth]{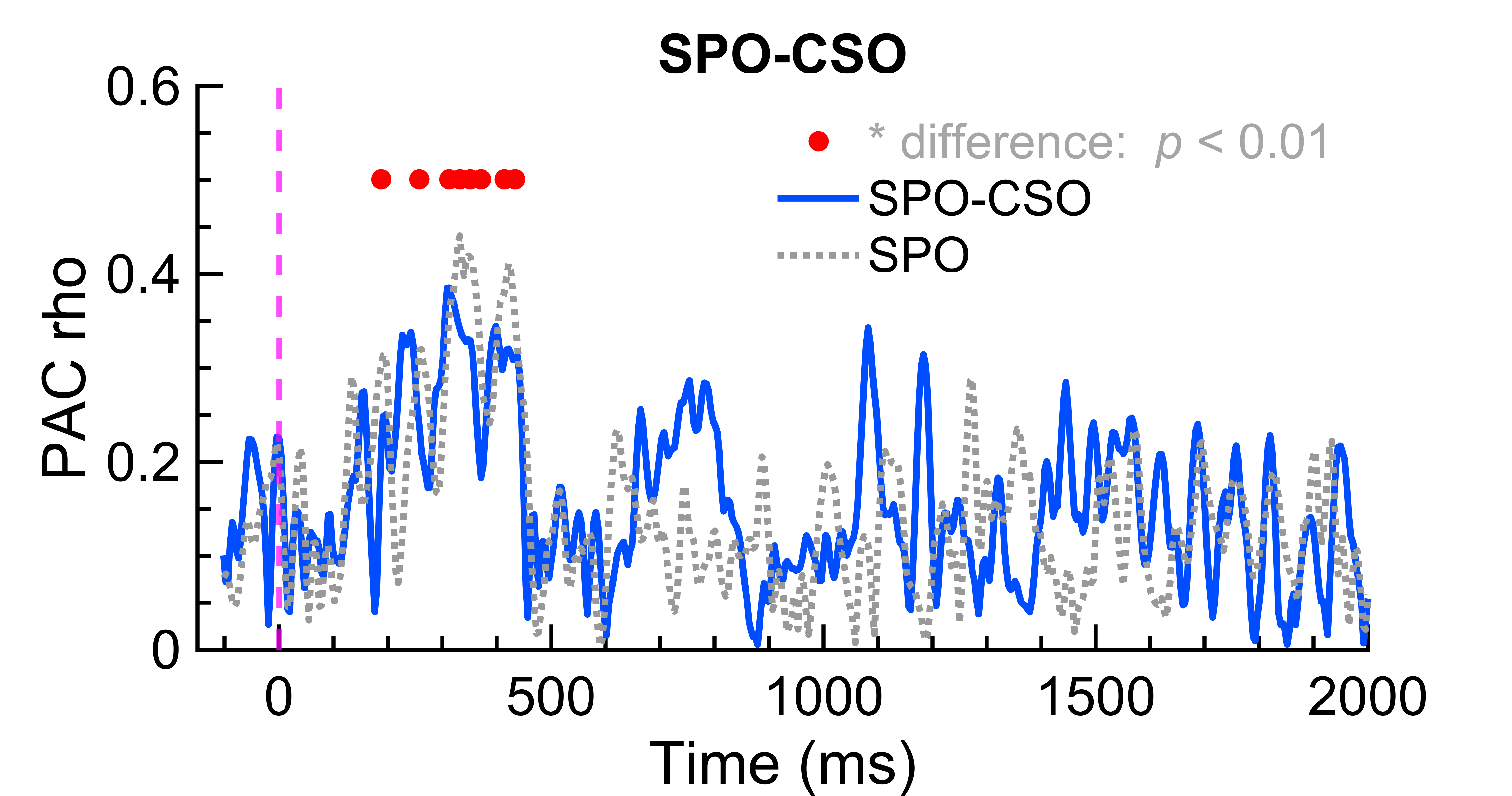}
  \caption{}
  \end{subfigure}
  \caption{Speech perception $\theta$ phase predicts covert speech low $\gamma$ amplitude between 200-500ms. PACs were calculated across trials and per sample of the time series after a Hilbert transform using the ERPAC toolbox \citep{Voytek2013}. Phase information was extracted only on SP classes and amplitude information was extracted only on the corresponding CS class (a,b). Statistical significance was calculated through surrogate testing. a and b show that SP's $\theta$ phase is significantly 'pseudo-coupled' to CS's low $\gamma$ amplitude compared to surrogate PACs between 200-500ms, with a similar coupling morphology to SP PAC. Red dotted lines indicate significance difference (\textit{p} <0.01) for blue lines.}
  \label{fig:xpac}
 \end{figure*}
  
\begin{figure*}[htbp!]
  \centering
  \begin{subfigure}[b]{0.475\linewidth}
    \centering

  \includegraphics[width=\linewidth]{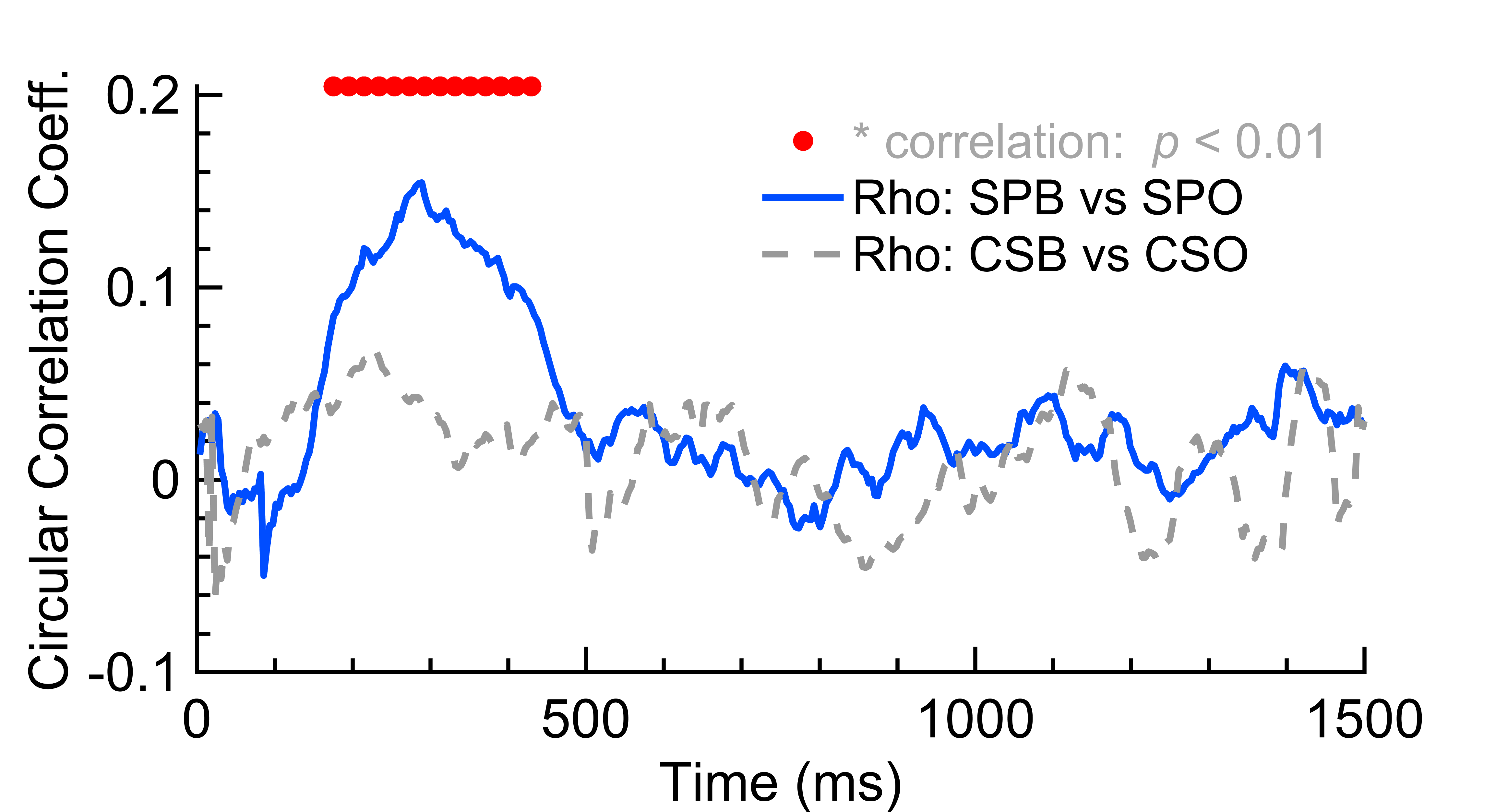}
  \caption{}
  \end{subfigure}
  \begin{subfigure}[b]{0.475\linewidth}
  \centering
  \includegraphics[width=\linewidth]{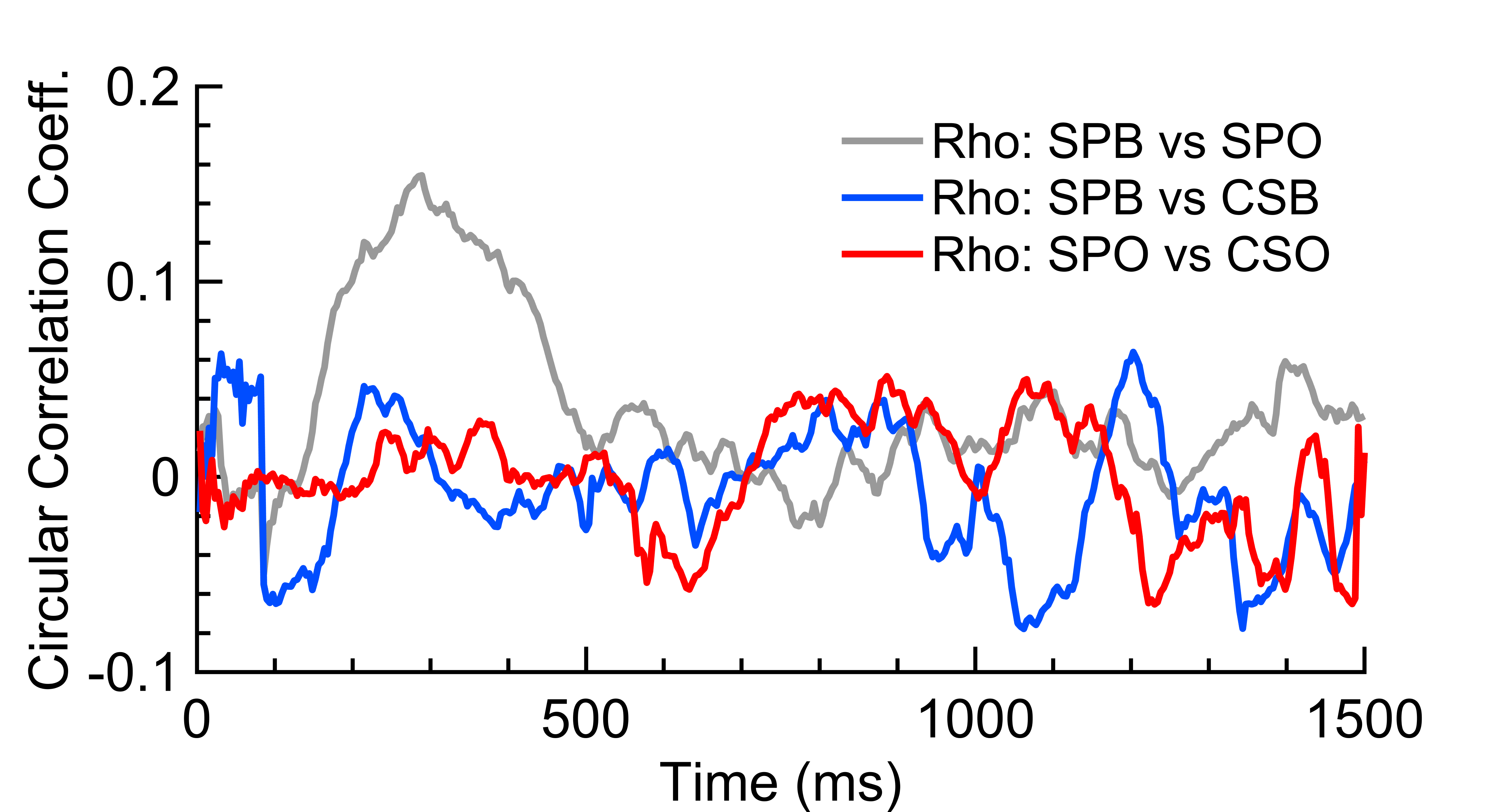}
  \caption{}
  \end{subfigure}

  \caption{$\theta$ activity is task dependent and serves different functions in speech perception and covert speech. $\theta$ phase of each signal was determined by calculating the angle of the Hilbert transform of the signal after a Butterworth band pass filter between 4-7Hz. The circular mean of the phase angles were calculated across all channels. Subsequently, the circular correlation between SPB and SPO $\theta$ phase was calculated for each time point and across trials. Data were pooled across all participants. $\theta$ phase was correlated in the two SP classes (a). Significant (\textit{p}<0.01) correlations were observed between 200-500ms for SPB-SPO, whereas CSB-CSO produced no significant phase correlations. Furthermore, no relationship was observed between the theta phases of SP and CS (b). Grey line depicts significant circular correlations of theta phase between SPB and SPO. Red and blue lines depict the SP-CS pairs which produced no significant $\theta$ phase correlations.}
  \label{fig:thetacorr}
  \end{figure*}

Considering the putative coordination between $\theta$ and $\gamma$ band activity \citep{Giraud2012,Hyafil2015}, PAC between these frequency bands were assessed across all classes (Fig. \ref{fig:pac}). Interestingly, single-channel PACs were observed specifically in only the right hemisphere temporal channels. Namely, in channel FT10, the PAC of both SP classes significantly departed from the surrogate PAC between 200-500ms  (\textit{p}<0.01), confirming that SP $\gamma$ amplitude produces a rhythm in keeping with the cadence of fluctuating $\theta$ phase. CS and rest were found to suppress or lack such a stable PAC, producing sparse distributions of significant departures from surrogates.
  
\begin{figure*}[htbp!]
  \centering
  
  \includegraphics[width=0.6\linewidth]{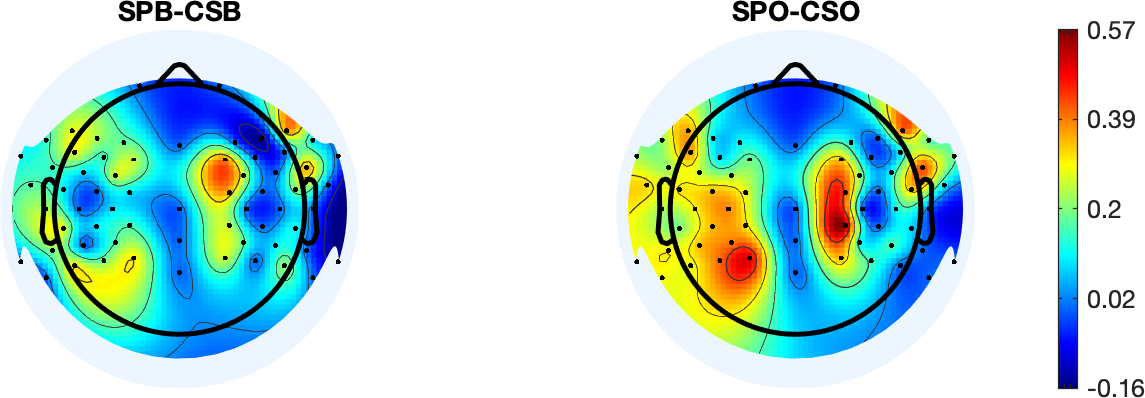}

   \caption{Topography of cross-trial $\gamma$ power correlations between corresponding classes. Data were pooled across all participants and sessions. $\gamma$ amplitude was extracted first through a Butterworth filter band-passed between 30-60Hz on preprocessed signals and then taking the absolute value of the Hilbert transform of said signals. Cross-trial $\gamma$ power was calculated by squaring the amplitude at each point in time, summing across trials, and normalizing by number of trials to result in a single time vector of power. s the Pearson's correlation coefficient was extracted for each channel between two opposing classes (e.g. SPB-CSB). High values (bright regions) correspond to high $\gamma$ power correlation across tasks (Pearson's rho). Only correlations with \textit{p}<0.05 are depicted.}
    \label{fig:gammacorr}
\end{figure*}

However, it was possible that CS $\theta$ activity may have served a separate function unlike that putatively observed in SP \citep{Restle2012,Albouy2017}, especially in the absence of salient stimulation. Therefore, the PAC between SP's $\theta$ and CS's $\gamma$ was assessed to determine whether the $\gamma$ amplitude of CS contains a task-specific rhythm (Fig. \ref{fig:xpac}a, b). Significant pseudo-PAC (\textit{p}<0.01) occurred again between 200-500ms for both words, confirming that CS's $\gamma$-band produced a rhythmic fluctuation specific to the time course of SP's $\theta$ synchronization (Fig. \ref{fig:ersp}a). Furthermore, relatively less but nevertheless significant pseudo-PACs were observed across words (e.g SPB-CSO), meaning that the SP $\theta$-CS $\gamma$ relationship contained both general and specific portions. This lack of specificity between SP $\theta$ and CS $\gamma$ was found to be due to significantly correlated $\theta$ phase patterns in SP of the two words between 200-500ms (circular correlation test \textit{p}<0.01) (Fig. \ref{fig:thetacorr}a). On the other hand, CS did not produce correlative $\theta$ phase patterns across words (Fig. \ref{fig:thetacorr}a) and across tasks (Fig. \ref{fig:thetacorr}b).

However, this did not portend that the $\gamma$-band responses of CS and SP words were general, as significant cross-trial $\gamma$ power correlations (Pearson correlation \textit{p}<0.05) were observed across tasks. Such correlations were observed in left temporal and temporo-parietal regions, the right fronto-temporal edge, and along the right motor to somatosensory regions (Fig. \ref{fig:gammacorr}), consistent to that seen in the topography of frequencies (Fig. \ref{fig:fttopo}). Importantly, significant $\gamma$-band power correlations were observed in channel FT10, where SP $\theta$-$\gamma$ and SP $\theta$-CS $\gamma$ PACs were observed.

\section{Discussion}

Using a $\mathtt{t}$-CWT method, the present study confirmed the hypothesis that CS largely utilizes higher frequency oscillations relative to SP. Crucially, we conclude that the $\gamma$-band likely functions similarly across tasks. Like SP, the $\gamma$ activity of CS was found to contain a processing rhythm time-locked to the cadence of event-related SP $\theta$ phase. Specifically, we suggest that CS's $\gamma$ activity likely depicts a phonological processing function similar to SP reported here and in previous studies \citep{Giraud2012,Pasley2012,Chang2010,Chang2016}. However, the lack of $\theta$-$\gamma$ PAC within CS suggests that CS's $\theta$ activity likely serves alternative roles to syllabic chunking seen in SP, possibly in preparatory motor or memory-related activity \citep{Restle2012,Albouy2017}. The present study represents the first investigation into the differences in oscillatory engagements between CS and SP and reports a relationship between SP's $\theta$ and CS's $\gamma$ activity. The findings reported here could enable the development of CS models based on SP signals, which can be used to train a CS BCI based on the passive perception of speech.

\subsection{Oscillations in speech perception and covert speech}

The main goal of the study was to assess the relative contribution and roles of major oscillatory dynamics in CS relative to SP. Thus, the oscillatory references of SP will be discussed first. A commonly synthesized interpretation from studies on SP investigating the role of oscillations is that each frequency band contributes to a dynamic sampling of speech items at varying temporal scales, referred to as multiplexing \citep{Gross2013}. The greater the frequency, the greater the resolution and detail at which a speech item is sampled. In the present study, we have identified that SP indeed utilizes the $\delta$, $\theta$, $\beta$, and $\gamma$ frequency bands, with significantly more prominent distinctions in the $\delta$ and $\theta$ bands than CS. This prominence can be attributed to the importance of tracking the speech envelope of salient percepts \citep{Luo2007,Giraud2012} and syllabic chunking \citep{Ghitza2012,Ghitza2013,Doelling2014}, features which seem to be common across languages \citep{Ding2017,Varnet2017}. Moreover, $\theta$ activity may produce generalized patterns during SP, as the frequency topography showed widespread and consistent distinctions across participants in the temporal and temporo-parietal channels.


The relatively high count of $\gamma$-band distinctions suggests that the driving force of delineating words in SP lies primarily in phonological processes \citep{Chang2010}. These activations were found focally in left and right temporal and temporo-parietal regions, which may loosely correspond to phonological regions of interests as reported through intracranial recording studies
\citep{Chang2010,Chang2016,Pasley2012}. Moreover, the significant $\theta$-$\gamma$ PAC between 200-500ms likely depicts a linkage between syllabic and phonological processes \citep{Giraud2012}, suggesting that the two oscillations may be inherently coordinated to focus $\gamma$ activity to specific times within syllabic time references upheld by $\theta$ \citep{Hyafil2015}. 

The timing of this PAC was likely linked to strong $\theta$-band synchronization occurring in this time period, as observed through ERSP. Interestingly, $\beta$-band synchronization followed shortly after. The $\beta$-band, producing a comparable amount of distinctions as $\gamma$, has been proposed to play a role in binding of semantic \citep{Weiss2003} and synactic information \citep{Bastiaansen2010}, being sensitive to the temporal alignment of ongoing speech \citep{Rimmele2018}. The observation of enhanced $\beta$-synchronization succeeding the $\theta$-$\gamma$ PAC can suggest that coordinated phonological and syllabic sampling becomes bound/conjoined into whole-word percepts via $\beta$ activity during SP.

CS, on the other hand, did not exhibit a diverse multiplexing relationship akin to SP, but rather favoured high frequency activity, namely $\beta$ and $\gamma$, for the distinction of words. The low count and sparse distribution of $\theta$ activity was likely due to lack of salient percepts in CS, suggesting that, unlike SP, $\theta$ activity in CS is focal and may serve a dissimilar function (discussed in Section 5.3). On the other hand, the high amount of distinctions in the $\gamma$-band suggested that this frequency is highly specific to word identity. Indeed, studies of overt and covert phoneme/word repetition tasks have shown differential $\gamma$-band augmentations in the temporal and temporo-parietal lobe \citep{Fukuda2010,Pei2011,Toyoda2014}, which may \textit{loosely} correspond to the frequency topography reported here. What the frequency topography does firmly suggest is that the widespread and high count of $\beta$- and $\gamma$-band distinctions likely indicated a greater degree of inter-participant consistency and can, in turn, allude to the existence of generalized activation profiles in temporal and temporo-parietal channels during CS. 

It is possible that the $\gamma$-band activity during CS corresponded to a corollary discharge, as enhanced fronto-temporal $\gamma$ synchrony has been observed during speech production tasks relative to perception conditions \citep{Chen2011,Ford2005}. Furthermore, in investigations of auditory verbal hallucinations (AVH), fronto-temporal $\gamma$ synchrony was found to be significantly suppressed in schizophrenic individuals \citep{Uhlhaas2006,Uhlhaas2010,Gallinat2004a}, suggesting an improper transmission of corollary discharge leads to phantom perceptions  \citep{Mathalon2008,Lutterveld2011}. Although such inter-reginoal synchrony was not investigated here, the fact that a suppression of this auditory prediction produces phantom perceptions of internal thoughts suggests that the pattern of activity in corollary discharge reflects the potentials during SP of the same words. This invites the hypothesis that CS's $\gamma$ activity serves a 'mirrored phonological' function to SP's $\gamma$ activity (discussed in Section 5.2). Similarly, it can follow that the observed $\beta$-band synchronization (occurring at the same time as SP) also signifies the temporal binding routines of phonological speech units enacted by the $\gamma$-band (Section 5.4).

\subsection{Common role of Gamma activity}

It was observed that CS and SP engage their oscillations differentially during speech processing. Interestingly, $\gamma$ distinctions were found to be the highest within tasks. As previously mentioned, a multitude of studies describe a phonological processing function of $\gamma$ activity during SP \citep{Chang2010,Chang2016,Pasley2012}. Although the methods of the present study were not sensitive to determining phonological cognitive load, we specifically asked whether CS's $\gamma$ activity contained a processing rhythm specific to $\theta$ activity. SP's $\gamma$ activity has been shown to keep a rhythm with respect to the rise and fall (periodicity) of its $\theta$ phase, which, when coupled, putatively allows individual phonemes to be processed in the context of larger syllabic units \citep{Giraud2012,Hyafil2015}. The existence of such a rhythm specific to a time frame would lend support, but not proof, for the hypothesis that $\gamma$ activity in CS serves a similar function to that in SP. Therefore, confirming this $\gamma$ rhythm would resolve an important step toward modelling CS from SP through a demonstration of possible functional equivalence.

However, a $\theta$-$\gamma$ PAC was not observed within the CS condition. This was likely due to $\theta$ activity in speech production (and its variants) being responsible for non-linguistic portions of the task such as motor \citep{Restle2012} and memory-related processes \citep{Albouy2017}. Indeed, task-dependent $\theta$ phase correlations were not observed within CS (Fig. \ref{fig:thetacorr}a). Thus, we asked if CS $\gamma$ would produce a rhythm that corresponds to the task- event-related cadence of SP's $\theta$ activity. Indeed, the results of the current study support this hypothesis with significant SP $\theta$-CS $\gamma$ PAC also occurring at 200-500ms. This coupling was found to be temporally sensitive and specific to this particular time period, as an otherwise random relationship would portend a sparsely distributed coupling pattern. This result confirms that CS's $\gamma$ activity is rhythmic, and importantly, specific to the periodicity of SP's $\theta$ phase, which putatively tracks syllabic quantities through the stimulus envelope \citep{Luo2007}. Therefore, the observed cross-task pseudo-coupling supports the idea that the $\gamma$-bands of SP and CS served similar functions.

The notion that this observed time-localized rhythmicity of CS $\gamma$ activity corresponds to a phonological processing function like SP may be \textit{loosely} entertained by the frequency topography, which shows that CS words are distinguished in the $\gamma$ band more consistently in the temporal and temporo-parietal channels. Intracranial studies employing overt and covert phoneme repetition tasks also report $\gamma$ activity in these regions, but specifically in the superior temporal gyrus and supramarginal gyrus \citep{Fukuda2010,Toyoda2014}, regions which have previously shown to play a role in phonological processing through fMRI investigations \citep{Okada2006,VandeVen2009,Venezia2016}. Although it is tempting to connect the present topographical results to the source localization in these studies, a word of caution is warranted as EEG is known to have poor spatial resolution and activations portrayed by the scalp map may not project ideally to the putative sources of speech processing.  

In contrast to the loose functional correspondence depicted by the topographical results, the finding of $\gamma$ rhythmicity in SP and CS in the same time frame (200-500ms) substantiates the interpretation that $\gamma$ activity served a similar function across tasks. The temporally co-localized $\gamma$ rhythms of SP and CS both seemed to correspond to transient $\theta$-band synchronizations in the same time period, likely caused by a phase resetting priorly. In SP, $\theta$ phase has been found to reset to the temporal edges of the speech envelope \citep{Gross2013} in order to initiate the coordination of processing at syllabic and phonemic levels \citep{Assaneo2018}. Similarly, $\theta$ phase has been found to reset to the onset of CS, resulting in strong phase-locking between 250-500ms that represents a temporal marker of CS processing \citep{Yao2020}. It remains inconclusive whether the present $\theta$-band synchronization in CS was a result of phase resetting \citep{Luo2007} or greater evoked potentials \citep{Obleser2012}. However, if like overt speech, CS tracks self-generated and temporally regular speech through neural oscillations, then the $\theta$ phase would necessarily reset to cause enhanced synchronization across trials \citep{Luo2007}. 

From a neural architectural perspective (i.e. neural circuits), such phase-resetting has been proposed to underlie information transmission such as communication through coherence \citep{Roberts2013} and the phase-dependent coordination of large scale neural networks for encoding and decoding during attention and goal-directed behaviours \citep{Canavier2015,Voloh2016}. It is thought that phase alignment through resetting forms predictable windows for integration which aids the coordinated parsing of segments \citep{Fries2009}. Hence, the transient $\theta$-band synchronizations observed here likely demarcated the points of processing in the tasks, indicating that CS and SP process words at the same time. Naturally, it then follows that the common occurrences of $\gamma$ rhythms, both of which are modulated and pseudo-modulated by SP's $\theta$ phase, represented a similar function between SP and CS. Since CS is a variant of speech production (only lacking overt articulation and production of sounds), it must generate a timely internal auditory prediction to match the processing of self-generated speech sounds \citep{Jack2019,Scott2013}. Therefore, under the view that CS is equivalent to self-generated SP without feedback, we propose that CS's $\gamma$-band response may have represented a similar function, potentially relating to phonological processing.

While the above discussion supports the idea that the $\gamma$-bands subserved similar functions across tasks, it may be further reasoned that $\gamma$ activity during CS represents a 'mirrored phonological' activation pattern, as previously pondered. This hypothesis emerged out of studies of AVH whereby an aberrant corollary discharge (i.e. $\gamma$ synchrony) results in phantom perceptions \citep{Mathalon2008,Lutterveld2011,Ford2005}. As the purpose of corollary discharge is to cancel out self-generated sounds, it follows that CS's $\gamma$-band response must continuously predict the sound patterns of ongoing speech. This invites the hypothesis that CS's $\gamma$ activity may produce similar activation patterns to that of SP's $\gamma$-band response, upon some transformation. As the current study did not analyze any existing correlations between the $\gamma$ activities, future studies are directed to employ distance correlation measures to quantify the predictability and/or dependence between the $\gamma$-band amplitudes of SP and CS.

However, some caution is warranted to the above interpretations as the current study involved only two speech tokens. Hence, future studies are directed to design studies with more diverse arrays of speech tokens for understanding the relationship between SP and CS's $\theta$- and $\gamma$-band responses. 
\subsection{Differential role of Theta activity}

It was found that SP and CS produced rhythmic fluctuations of $\gamma$-band activity that correlated to the tracking of the speech envelope by SP's $\theta$ activity. However, the lack of $\theta$-$\gamma$ PAC occurring \textit{within} CS suggested that the $\gamma$ band response of CS retained its own processing rhythm in the absence of modulation by its $\theta$ phase. This observation is corroborated by a lack of $\theta$ phase correlations occurring between SP and CS (Fig. \ref{fig:thetacorr}b). Thus, it is possible that $\theta$ activity served a dissimilar function to syllabic chunking as seen in SP, or that the relationship between $\theta$ and $\gamma$ in CS cannot be described by a coupling of phase and amplitude. Similar to the current results, \citeauthor{Hermes2014} (\citeyear{Hermes2014}) showed that $\theta$-$\gamma$ PAC is suppressed during CS in Broca's area, the temporo-parietal junction, and middle temporal gyrus, and that its $\theta$ power is anti-correlated to high frequency power. Perhaps more counter-intuitively, $\theta$-$\gamma$ PAC has been reported to increase in patients during AVH as thoughts manifest as phantom perception \citep{Koutsoukos2013}; the inverse of which suggests that a normal thought would suppress this $\theta$-$\gamma$ coupling. These results beg the question: if CS's $\theta$ band synchronizes in the same time period (Fig. \ref{fig:ersp}b) as the emergence of its $\gamma$ rhythm, what is the role of $\theta$ with respect to $\gamma$ in CS?

Considering that stimulation of a major dorsal stream area (posterior inferior frontal gyrus) with $\theta$ burst stimulation facilitates speech repetition accuracy \citep{Restle2012}, it may be reasoned that $\theta$ activity during CS correlates to motor planning and activity. Indeed, 4-7Hz also corresponds to the mandibular movement rate during articulation \citep{Giraud2007}, which is also demonstrated by an enhanced coupling between motor and auditory areas during syllable presentation at 4.5Hz \citep{Assaneo2018,Poeppel2020}. These studies indicate that the $\theta$-band represents a preferred articulatory rhythm and thus has motor origins in speech production. If this interpretation holds, it would suggest that phonological and articulatory processing in CS are independent in the context of PAC, but potentially related by some other measure. For instance, it may be possible that $\theta$-based articulatory expressions induces $\gamma$-based corollary discharge that pre-contains the sensory predictions outlined by the motor code and rules out the need for coupling between the two frequency bands. Indeed, $\theta$ coherence \citep{Ford2002} and $\gamma$ synchrony \citep{Uhlhaas2006,Uhlhaas2010,Gallinat2004a} has both been found to be significantly reduced in schizophrenic patients with AVH, suggesting that the independent suppression of synchronization in the two oscillations each plays a role in an aberrant corollary discharge mechanism. Ameliorating the current results with the findings of these studies can lead to the the hypothesis that $\theta$-based motor discharges may inform $\gamma$ activity in CS, but not be linked through a PAC.


Alternatively, or perhaps in parallel, $\theta$ oscillations in the dorsal stream may also work by enhancing auditory working memory \citep{Albouy2017}, potentially in the form of access to lexical stores \citep{Piai2014}. This is consistent with the model proposed by \citeauthor{Indefrey2004} (\citeyear{Indefrey2004}) where word production has been suggested to initiate with the lexical concept. More empirically, in a series of studies investigating oscillatory power during covert word reading \citep{Bastiaansen2005} and lexical deicsion-making \citep{Bastiaansen2008}, $\theta$ power - or local synchrony - was found to be modulated as a function of lexicality, peaking between 300-500ms. Thus, $\theta$ activity as accessing the mental lexicon is a sensible interpretation, as instantiating a lexical memory can initiate subsequent unitary/phonological processing by the $\gamma$-band, but not be necessarily linked through phase and amplitude. Indeed, $\gamma$-band activity has been suggested to be necessary for the formation of both phonological and lexico-semantic representations of words through repetition and homophone priming tasks \citep{Matsumoto2008}. It is thus possible that $\theta$ activity registers the broader lexical framework for the auditory prediction and informs $\gamma$-band corollary discharge, which seemingly provides the sensory/phonological representation of the word. Although the singular role of $\theta$ activity in speech production is being debated, considering that slow oscillations can synchronize between widely distributed brain areas \citep{Buzsaki2004}, the different role of $\theta$ activity in CS may encompass motor-related activity and access to lexical memory simultaneously or in a cascading manner. However, the methods of the present study were not sensitive to understanding the motor or lexical load by CS's $\theta$ activity. 

\subsection{General discussion and Limitations}

The processing of CS brought about a comparable amount of distinctions in the $\beta$-band as the $\gamma$-band. While $\beta$ activity for speech production has been reported to play a role in motor activity and motor preparation \citep{Mersov2016,Piai2015}, in language-related processes, it has importantly been proposed to serve as a top-down modulatory signal for the temporal management of ongoing speech \citep{Rimmele2018}. Indeed, in language processing, $\beta$ activity has been shown to play a role in the timely binding of synactic \citep{Bastiaansen2010} and semantic \citep{Weiss2003} information. More broadly, \citeauthor{Weiss2012} (\citeyear{Weiss2012}) assert that synchronized $\beta$-band oscillations serve to bind the contents of distributed set of neuronal populations into one coherent memory unit. The role of $\beta$ activity in temporal binding routines is supported by studies revealing significant $\delta$-$\beta$ PAC \citep{Arnal2015,Keitel2017,Keitel2018,Morillon2019}, for the registration of words, phrases, and sentences by $\delta$ must necessarily emerge from the temporal bindings along the hierarchy of speech units (phonemes to words). Although the current study did not investigate this coupling, the observation of temporally co-localized $\beta$-band synchronizations between CS and SP, succeeding the common $\theta$-band synchronization (discussed to signify the processing of SP and CS), invites the hypothesis that the $\beta$-band, too, served a similar function across tasks, namely in enacting binding routines of phonological/syllabic items into broader whole-word percepts.

Finally, it should be noted that the phase-amplitude relationships between SP's $\theta$ and CS's $\gamma$ were found to be general as weaker, but nevertheless existent, pseudo-couplings were observed also across words (e.g. SPB $\theta$-CSO $\gamma$). However, this does not necessarily portend that the $\gamma$ activities themselves are general, as no gamma-band correlations were observed. Indeed, CS and SP both produced a significant amount of distinctions in the $\gamma$-band, suggesting that $\gamma$ activity is likely specific to words. Instead, general nature of SP-CS PAC can be attributed to the lack of diversity in the current lexicon, which varied only between 1-2 syllables and spoken at the same rate. This evidently led to non-divergent $\theta$ patterns: average $\theta$ phases were found to be significantly correlated between 200-500 in the two SP classes, making this frequency band less pertinent to distinction of SP words than $\gamma$ activity (Fig. \ref{fig:thetacorr}a). In contrast, studies using a larger vocabulary and sentential speech tokens have shown that $\theta$ phase adjusts to syllabic rate and the number of syllables \citep{Assaneo2018,Lizarazu2019,Ding2017,Ding2017a}. Therefore, future studies should experiment with a richer lexicon with more syllable counts, possibly embedded in sentential forms, in order to determine whether CS's $\gamma$ activity forms a specific relationship to the putative syllabic tracking by SP's $\theta$ activity. Such studies will assist in determining whether this pseudo-coupling between SP and CS simply reflects task demands or reflects common neurolinguistic processing. Furthermore, future studies are directed to determine whether correlations exist between the $\gamma$-band responses of SP and CS. These investigations should provide substantial supports for the hypothesis that CS $\gamma$ activity also reflects a phonological process.


\section{Conclusion}
The present study represents the first to investigate the similarities and differences with respect to oscillatory engagement during CS and SP. We found that CS favours higher frequency activity that likely reflects corollary discharge. Specifically, we found that CS's $\gamma$-band response has a similar rhythmic pattern to SP, possibly representing a similar phonological process. These findings substantiate the results of \citeauthor{Oppenheim2008} (\citeyear{Oppenheim2008}) who describe CS to contain robust phonological information, and further suggests that CS and SP may share a common function when it comes to $\gamma$ activity. Contrarily, we assert that $\theta$ activity in CS and SP play different roles possibly via differential processing through dorsal and ventral streams, respectively. Understanding the relative oscillatory engagements and their functional correlates in the two tasks are elemental to the modelling of CS based on SP signals. Therefore, the present work can lead to the development of CS BCIs through the passive perception of speech, which can help hurdle the difficulties of training by rendering the training process passive. In order to achieve this modelling, we direct future studies to investigate the details of the relationship between SP's $\theta$ activity and CS's $\gamma$ activity, as well as similarities in the $\gamma$-band responses for further confirmation of a common function of $\gamma$ activity.

\section{Acknowledgements}
We thank Christine Horner and Sarah Holman for their help in data collection, and Ka Lun Tam and Pierre Duez for assisting in developing the protocol.

\bibliographystyle{apalike}  
\bibliography{library}  

\end{document}